\documentclass[acmsmall]{acmart}

\usepackage[english]{babel}
\usepackage{blindtext}
\renewcommand\footnotetextcopyrightpermission[1]{} 
\setcopyright{none}
\acmDOI{}
\acmISBN{}

\usepackage{graphicx} 

\usepackage{soul}
\usepackage{tikz}
\usepackage{amsmath}
\usepackage{titlesec}
\usepackage{xspace}
\usepackage{enumitem}
\usepackage{subfigure}
\usepackage{mathtools}
\usepackage{xurl}
\usepackage[utf8]{inputenc}

\usepackage{soul}
\usepackage{appendix}
\usepackage{float}

\usepackage{pifont}
\usepackage{multicol,multirow}
\usepackage{tablefootnote}
\usepackage{capt-of}
\usepackage{caption} 
\usepackage{sidecap}
\usepackage[normalem]{ulem}
\usepackage{graphicx,wrapfig,lipsum}

\newcommand{\ie}{\emph{i.e.,}\xspace}

\newcommand{\name}{{\sc Titan-Next}\xspace}
\newcommand{\msteams}{{\sc Microsoft Teams}\xspace}
\newcommand{\teams}{{\sc Teams}\xspace}
\newcommand{\tilda}{{\sc Titan}\xspace}
\newcommand{\azure}{{Azure}\xspace}
\newcommand{\rohan}{\textcolor{black}}
\newcommand{\db}{\textcolor{black}}

\makeatletter
\newcommand\footnoteref[1]{\protected@xdef\@thefnmark{\ref{#1}}\@footnotemark}
\makeatother

\titlespacing\section{2pt}{2pt plus 1pt minus 1pt}{2pt plus 2pt minus 2pt}
\titlespacing\subsection{2pt}{2pt plus 1pt minus 1pt}{2pt plus 2pt minus 2pt}
\titlespacing\subsubsection{2pt}{2pt plus 1pt minus 1pt}{2pt plus 2pt minus 2pt}
\begin{document}

\title{Saving Private WAN: Using Internet Paths to Offload WAN Traffic in Conferencing Services}

\author{Bhaskar Kataria}
\email{bk478@cornell.edu}
\affiliation{%
  \institution{Microsoft Research}
  \country {India}
}

\author{Palak LNU}
\email {palak7372@gmail.com}
\affiliation{
  \institution{Microsoft Research}
  \country {India}
}

\author{Rahul Bothra}
\email {bothra2@illinois.edu}
\affiliation{
  \institution{UIUC}
  \country{USA}
}

\author{Rohan Gandhi}
\email {rohangandhi@microsoft.com}
\affiliation{
  \institution{Microsoft Research}
  \country {India}
}

\author{Debopam Bhattacherjee}
\email {debopamb@microsoft.com}
\affiliation{
  \institution{Microsoft Research}
  \country {India}
}

\author{Venkata N. Padmanabhan}
\email {padmanab@microsoft.com}
\affiliation{
  \institution{Microsoft Research}
  \country {India}
}

\author{Irena Atov}
\email {iratov@microsoft.com}
\affiliation{
  \institution{Microsoft}
  \country{USA}
}

\author{Sriraam Ramakrishnan}
\email {sriraamr@microsoft.com}
\affiliation{
  \institution{Microsoft}
  \country{USA}
}

\author{Somesh Chaturmohta}
\email {someshch@microsoft.com}
\affiliation{
  \institution{Microsoft}
  \country{USA}
}

\author{Chakri Kotipalli}
\email {chakoti@microsoft.com}
\affiliation{
  \institution{Microsoft}
  \country{USA}
}

\author{Rui Liang}
\email {rui.liang@microsoft.com}
\affiliation{
  \institution{Microsoft}
  \country{USA}
}

\author{Ken Sueda}
\email {ksueda@exchange.microsoft.com}
\affiliation{
  \institution{Microsoft}
  \country{USA}
}

\author{Xin He}
\email {xhe@microsoft.com}
\affiliation{
  \institution{Microsoft}
  \country{USA}
}

\author {Kevin Hinton}
\email {kehin@microsoft.com}
\affiliation{
  \institution{Microsoft}
  \country{USA}
}
\renewcommand{\shortauthors}{Bhaskar Kataria, Palak, Rahul Bothra, Rohan Gandhi, Debopam Bhattacherjee, Venkata N. Padmanabhan et al.}

\date{}

\maketitle
\sloppy

\section*{Abstract}

\db{Large-scale video conferencing services incur significant network cost while serving surging global demands. Our work systematically explores the opportunity to offload a fraction of this traffic to the Internet, a cheaper routing option offered already by cloud providers, from WAN without drop in application performance. First, with a large-scale latency measurement study with $3.5$~million data points per day spanning $241K$ source cities and $21$ data centers across the globe, we demonstrate that Internet paths perform comparable to or better than the private WAN for parts of the world (e.g., Europe and North America). Next, we present \tilda, a live ($12+$ months) production system that carefully moves a fraction of the conferencing traffic to the Internet using the above observation. Finally, we propose \name\xspace -- a research prototype that jointly assigns the conferencing server and routing option (Internet or WAN) for individual calls. With $5$~weeks of production data, we show \name reduces the sum of peak bandwidth on WAN links that defines the operational network cost by up to $61\%$ compared to state-of-the-art baselines.} We will open-source parts of the measurement data.
 
\section{Introduction}
\label{sec:intro}

Conferencing services such as Zoom\cite{zoom:web}, Microsoft Teams\cite{teams:web}, DingTalk\cite{dingtalk:web}, and Google Meet\cite{meet:web} have become an indispensable part of \db{our society},  especially since the COVID-$19$ pandemic. However, the skyrocketing growth in demand\cite{teamsgrowth:web} has also resulted in higher costs incurred by such services\cite{xron:sigcomm23, switchboard:sigcomm23}. Usually, such large-scale conferencing services use dedicated Media Processor (MP) servers \cite{xron:sigcomm23, switchboard:sigcomm23}) in cloud data centers (DCs) that receive media streams (audio, video, and screen-share) from users, process, and redistribute them.
The cloud providers' private WANs (wide-area networks) have been the default choice~\cite{wan:infocom20, switchboard:sigcomm23} to carry traffic between users and the MP servers. 
\db{With such routing, the conferencing traffic ingresses into and egresses from the WAN closest to the user (not the MP server) thus consuming significant WAN resources and inflating the operational cost for the application.}
\db{Cloud providers have started providing the \textit{Internet routing} option \cite{gcp:internetprice:web, azure:internetprice:web} that allows application traffic to ingress/egress closer to the cloud (MP) server, thus reducing the WAN load.}
This Internet routing option is significantly cheaper than WAN (by up to $53\%$~\cite{azure:internetprice:web, gcp:internetprice:web}) \db{motivating us to explore if a fraction of the conferencing traffic could leverage this routing option, without affecting user experience, thus reducing operational network cost.}

First, we shed light on the performance of the Internet versus WAN routing from a hyperscalar. While both network loss and latency can affect user experience, the former is mitigated to some extent through application layer mechanisms such as \cite{tambur:nsdi23,bitag:latency2022} while the latter is harder to tackle since it is a more fundamental impediment to interactivity.
To understand if we can leverage Internet paths to reduce costs of conferencing services \db{without affecting application performance}, we measure latencies using the WAN and the Internet routing options. Our measurements are piggybacked on  \msteams (referred as \teams in rest of the paper) -- a large-scale conferencing service. We run $3.5$~million measurements on average per day for $12$ months from $241K$ cities (distinct population centers) across the globe to $21$ \azure{}\footnote{\label{pseudonym} \teams is hosted on \azure; so, we measure latency to \azure.}
DCs. Our measurements show that \textit{the Internet offers latency as good as WAN\footnote{In rest of the paper, we use WAN to denote private WAN of \azure, and Internet as public Internet.} or even better} in parts of the world -- especially in Europe (including UK) and North America. 
Such an observation is promising to move some of the WAN traffic to the Internet to reduce costs. We proceed in two phases: ($1$) We move a fraction of the WAN traffic to the Internet prioritizing \textit{safety} over optimality without changing MP DC assignments to calls. To that extent, we built \tilda that has been in production for the last $1$ year. ($2$) We \db{built \name as a research prototype that further reduces traffic on WAN through joint assignment of MP DCs and Internet routing}.

Next, we present \tilda\xspace -- a system that carefully moves a fraction of the total \teams traffic to the Internet. A key concern with Internet offload is that \teams consumes significant bandwidth, and moving all of its traffic to the Internet can result in network congestion and poor user experience. Measurements through \textit{ping} are too lightweight to gauge capacity on the Internet. \tilda moves traffic to the Internet iteratively. In each iteration, it increases traffic on the Internet and measures the latency, loss, jitter and other network and application metrics. The iterations terminate when a subset of the metrics indicates early signs of deteriorating performance. Also, We prioritize \textit{safety} as we stop moving traffic to the Internet after certain point even if there is no deterioration in performance. In this paper, we present the design of \tilda and our experiences with \db{moving large amounts of live \teams production traffic to the Internet (\S\ref{sec:intmigration}) with \tilda for the last $12${}$+$ months}.

Finally, we present \name. \tilda focuses on choice of routing given the selection of MP servers to calls is fixed. We do even better using \name by jointly optimizing the MP selection and routing based on the Internet capacity calculated by \tilda. \name is based on three key ideas: ($a$) The MP DC selection and the routing option need to be jointly optimized as the former has bearing on choosing latter. ($b$) Interactivity of the call depends on \textit{maximum end-to-end latency} between participants. Our results show that the participants are sensitive to this metric. Hence, it should be considered while making assignments. ($c$) \name reduces call migrations using \textit{reduced call configurations}, which is an abstraction for the resource needs of a call (\S\ref{sec:design:migration}). 

\name formulates a Linear Program (LP) that uses these key ideas to jointly determine the MP DCs and routing options for calls. \name is a research prototype that currently works in a \db{\textit{shadow mode} alongside \tilda (live in production) and computes the potential savings of doing assignments differently than \tilda.}
We evaluate \name using $5$ weeks ($4$ weeks of training data + $1$ week for evaluation) of call traces from production with $O$($10$~million) calls on a weekday. We compare \name against \tilda, Weighted Round Robin (WRR) and Locality First (LF) baselines similar to policies used in production. Our results show that: ($a$) \name can reduce the sum of peak bandwidth across WAN links by up to $61\%$, ($b$) \name achieves end-to-end latency close to LF which specifically optimizes for latency. ($c$) It can cut down the number of call migrations (details in \S\ref{sec:design:migration}) across DCs by up to $66\%$. 

In summary, we make three concrete contributions: ($a$) \textbf{Real measurements:} We present results comparing WAN vs. Internet latency using a large-scale measurement study. Our results show that the latter is comparable or even better in parts of the world. ($b$) \textbf{Production state-of-the-art:} We present our design of \tilda (motivated by results in $a$) for moving \teams calls to the Internet paths in \textit{quality-controlled} manner. We also share experiences in production.
 
($c$) \textbf{Research prototype:} We present \name that jointly assigns MP DC and routing option to reduce the WAN traffic (and costs) using prediction, offline computation plan, and real-time assignment. \name can cut down the sum of peak WAN traffic by up to $61\%$. 

We will open-source parts of the measurement data. This work does not raise any ethical issues.

\section{Background}
\label{sec:back}

\subsection{Primer on conferencing services}
\label{sec:back:call}

\teams and other large-scale conferencing services (CS) are known to host large volumes of calls globally. Each call is assigned a Media Processor (MP) server hosted on a cloud data center (DC) (called cloud region). Each participant in a call could generate up to $3$ distinct streams -- audio, video, and screen-share, which are sent to the cloud-hosted MP server, which in turn processes and forwards the streams to other participants in the call. For capacity, availability, and performance reasons, a large-scale CS usually has MP servers hosted in multiple DCs globally. 

\subsection{Problem: MP DC selection and routing}
\label{sec:back:mp}

The key problem in \teams is to determine the MP server for each call along with routing option for each participant, while balancing the user experience and costs. Such a problem has four aspects: (a) \textbf{Resource provisioning:} \teams is a \textit{first party CS} with access to compute and network resources from \azure. Due to its scale, high uptime requirements, \teams requires vast compute and network resources\footnote{MPs have no/minimal storage requirements. They primarily process and re-distribute streams.} that cannot be met through uncertainty of \textit{pay-as-you-go} models.  \teams rather provisions dedicated compute (MP) servers in advance. On the other hand, network bandwidth is more readily shareable with other \azure services. Therefore, though the network is provisioned in advance based on the anticipated need of all services including \teams, the billing is done based on the \textit{peak usage} of individual services.
(b) \textbf{MP DC selection:} based on the resources provisioned, \teams needs to assign MP DC for each call. Such an assignment needs to take into account the DC-wise compute provisioning, the geo-distribution of the call participants, and call experience. (c) \textbf{Route selection:} while \teams uses cloud provider's WAN, as we show in the next sections, the Internet provides performance comparable (or better) than WAN for some parts of the world. Thus, \teams needs to determine the routing option (WAN or Internet) for individual participants on a call. (d) \textbf{MP selection:} Once the MP DC is selected for a call, \teams needs to select the actual MP server in that DC to host the call.

Switchboard~\cite{switchboard:sigcomm23} efficiently addresses (a) and (b) assuming control of the amount of resources provisioned. In this work, we jointly do (b) and (c) assuming the resources are already provisioned. We use state-of-the-art load balancers for (d).

\textbf{Metrics of interest:} The key metrics of interest for \teams are: (a) User experience: user experience is sensitive to network latency, loss, and jitter\cite{mos:hotnets23}. \teams and conferencing services, in general, tackle jitter to a large extent using jitter buffers~\cite{bitag:latency2022}, and network loss (to certain extent) through error correction and error concealment~\cite{bitag:latency2022}. (b) Compute and network costs: Compute comprises the MP servers in DCs that are already paid for. \teams currently uses \azure's WAN where the network cost depends on the \textit{peak usage} (similar to \cite{pretium:sigcomm16, switchboard:sigcomm23}). \azure is one of the largest cloud provider, and \teams is among the top services by traffic rate in \azure. Thus, it is imperative to reduce network cost for \teams.

\subsection{Internet versus WAN routing}
\label{sec:back:internet}

\begin{figure*}
\centering
    \begin{minipage}{.48\textwidth}
        \centering
        \includegraphics[width=0.95\textwidth]{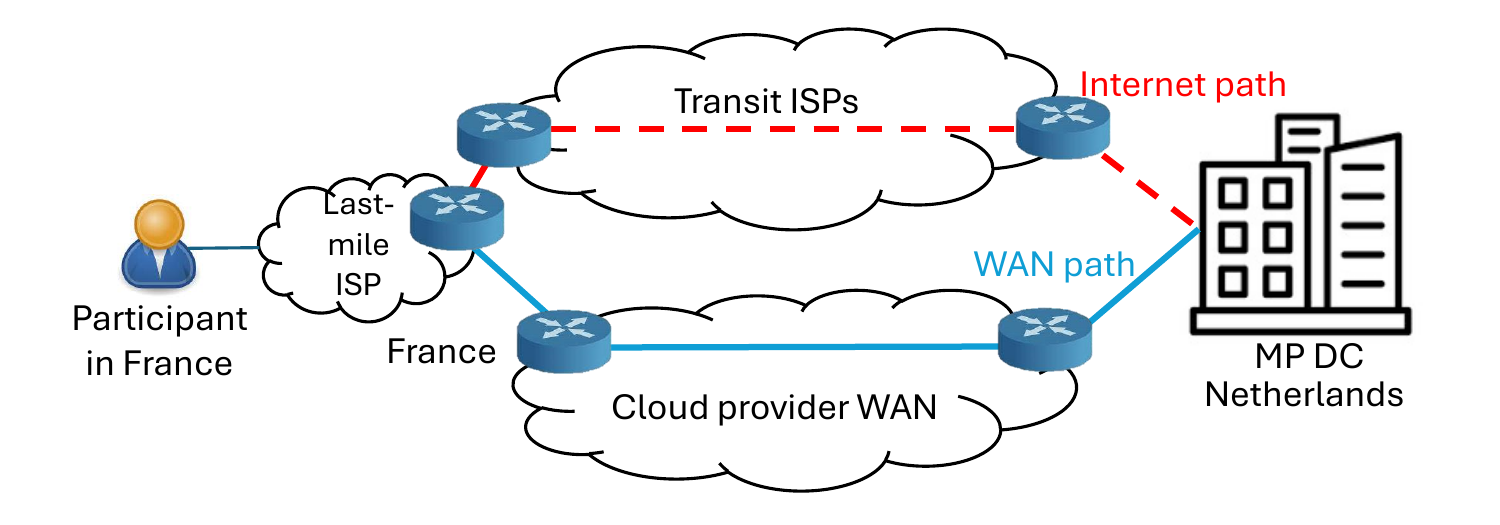}
        \caption{WAN versus Internet routing. 
        Using WAN routing, the traffic from MP exits the WAN closest to the user (cold-potato). Using Internet routing, the traffic exits the WAN closest to the DC (hot-potato).} 
        \label{fig:back:hot}
    \end{minipage}%
    \hspace{0.2cm}
    \begin{minipage}{0.48\textwidth}
        \centering
        \includegraphics[width=0.95\textwidth, page=1]{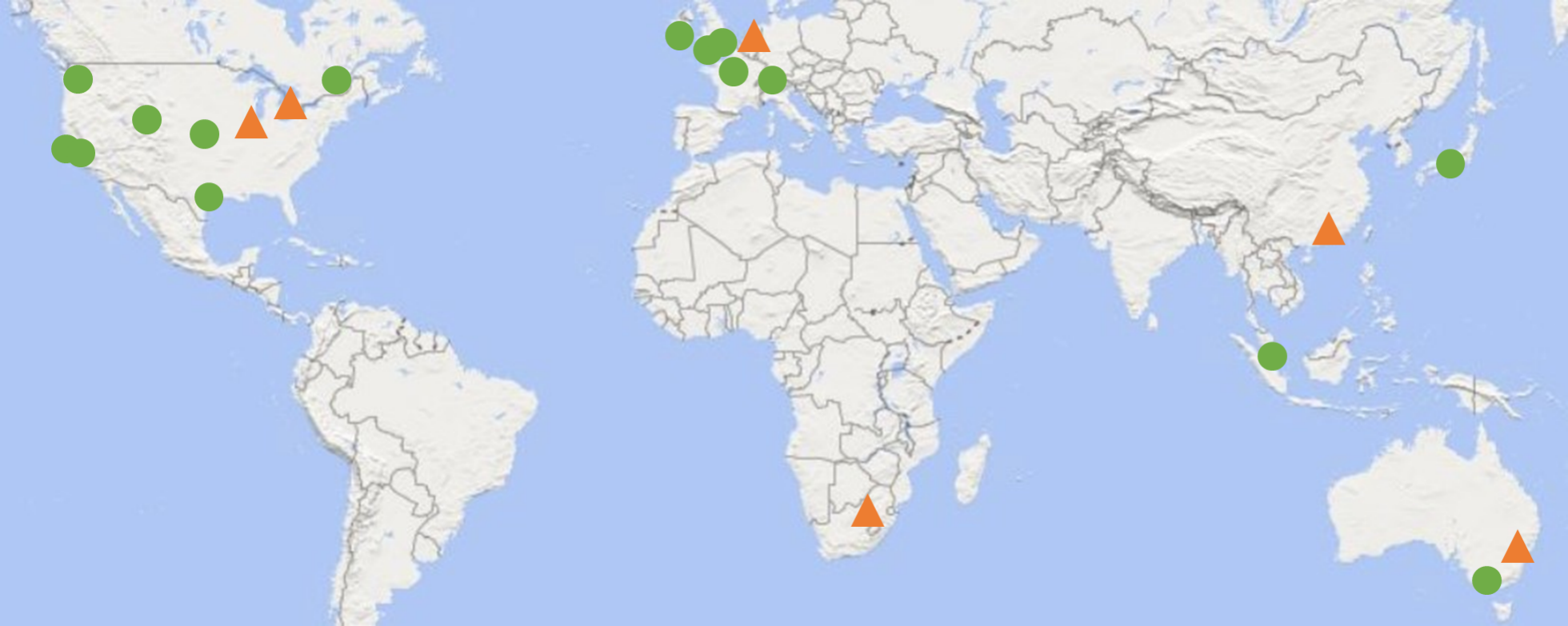}
        \caption{Locations of the $21$ \azure DCs used in the measurements. Orange triangles denote the locations of representative DCs used in Fig.\ref{fig:measure:latencyheatmap}.}
        \label{fig:measure:locations}
    \end{minipage}%
\end{figure*}

WAN has been a default choice to route the traffic between the cloud VMs and the users in \azure. Large cloud platforms including AWS, Azure, and GCP have started offering an alternate \textit{Internet routing option}, which customers/services could opt for. This is essentially hot-potato routing -- egress (likewise, ingress) traffic exits (enters) the Internet closest\footnote{Usually, multiple transit provider options; BGP picks one.} to the hosted service (Fig.\ref{fig:back:hot}). Internet paths are cheaper than WAN up to $53\%$\cite{azure:internetprice:web, gcp:internetprice:web}, e.g., GCP charges \$$0.15$ and \$$0.075$ for data transfers per GB\cite{gcp:internetprice:web} using WAN and Internet respectively for Singapore region. This observation motivates us to investigate \db{if a fraction of the \teams traffic can be moved to cheaper Internet paths while not compromising on the user experience. Additionally, as \teams peak traffic on the WAN is reduced, it provides more bandwidth for other services and reduces long-term capacity provisioning.}
Lastly, our study makes it evident that the Internet also provides a fall-back option to WAN. \db{We discuss in \S\ref{sec:intmigration:experiences} how Internet paths could augment WAN capacity during fiber cuts.}

\section{Internet paths good enough?}
\label{sec:internet}

We discuss measurement results contrasting the latency of the Internet and WAN paths.

\begin{wraptable}{l}{0.45\textwidth}
    \small
    \caption{
    Scale of our measurements.}
    \begin{tabular}{|c|c|c|}
        \hline
        \textbf{Geography} & \textbf{Unique values} \\
         \hline
        Avg. \#measurements/day & 3.5 million \\
        \hline
        Source country & 244 \\
        \hline
        Source city & 241,777 \\
         \hline
        Source ASN & 61,675 \\
         \hline
        IP subnets & 4,731,110 \\
        \hline
        Destination DCs & 21 \\
        \hline
    \end{tabular}
    \label{tab:internet:stats}
    \vspace{-0.1in}
\end{wraptable}

\textbf{Methodology:} We setup $42$ VMs (virtual machines; $2$ per DC) in $21$ \azure DCs (see Fig.\ref{fig:measure:locations}) across the globe. In each DC, one VM uses the Internet path and the other VM uses the WAN path. Both VMs host HTTPS servers that serve a $1${}$\times${}$1$ image (with some metadata) upon receiving requests from clients. A load-balancer assigns client requests to one of the $42$ VMs using round-robin scheduling. \teams has multiple $100$~million monthly active users. Our latency measurements span $12$~months (from June 2023 to June 2024) thus capturing $\sim${}$1.2$~billion measurements. The data is anonymized to remove any Personally Identifiable Information. For each test, the VM (location known) logs the timestamp of the test, $/24$ masked client IP address, and the request round-trip time (RTT). The client's IP address is translated offline to the client's country, city, and ASN using a proprietary geolocation database having high accuracy. The RTT takes into account only the GET request/response round-trip and disregards any HTTPS connection setup time. \textit{We refer RTT as "latency" for all further analyses}. Table \ref{tab:internet:stats} shows the statistics.

\begin{figure*}[t]
\centering
\subfigure[North America]
{
\includegraphics[width = 0.24\textwidth]{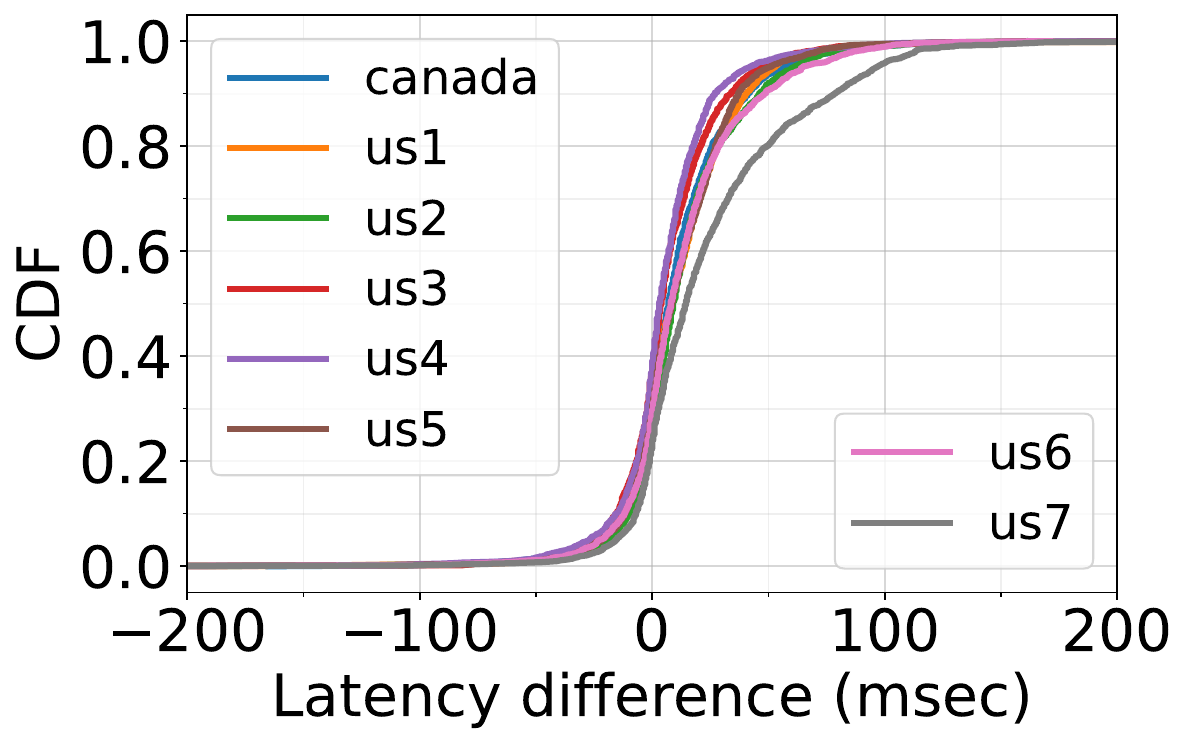}
\label{fig:measure:overall:con1}
}
\hspace{-0.3cm}
\subfigure[Europe]
{
\includegraphics[width = 0.24\textwidth]{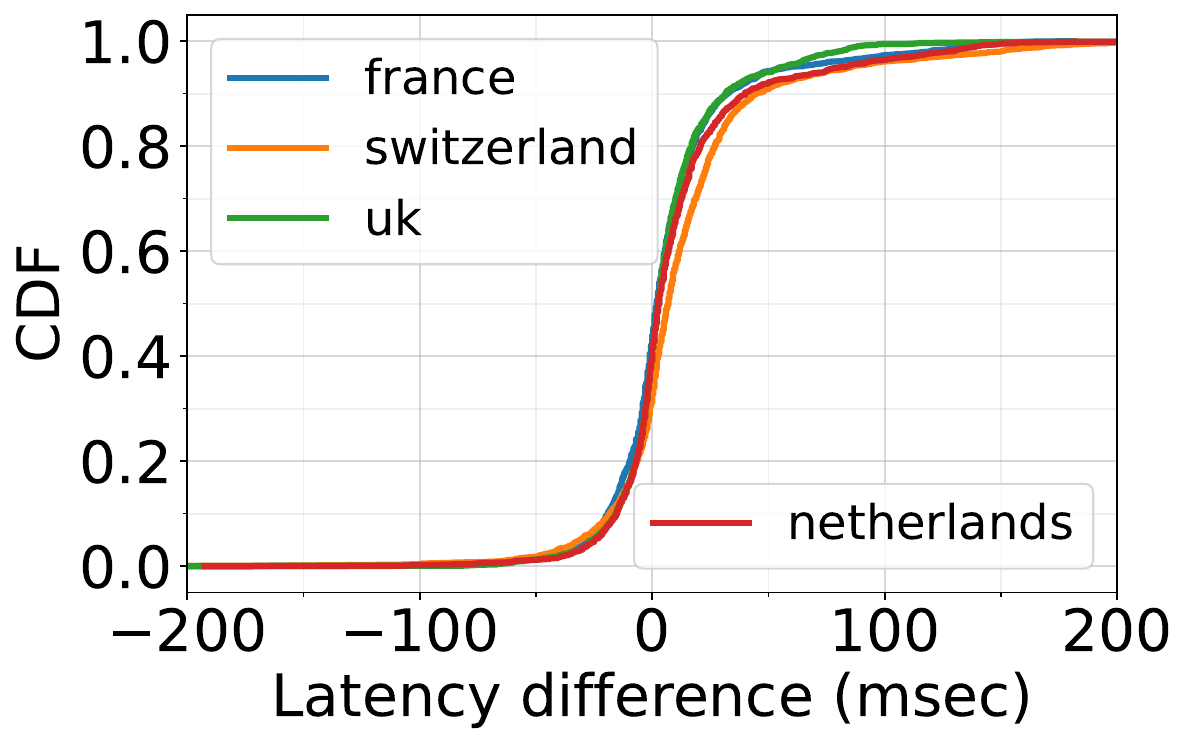}
\label{fig:measure:overall:con2}
}
\hspace{-0.3cm}
\subfigure[Asia]
{
\includegraphics[width = 0.24\textwidth]{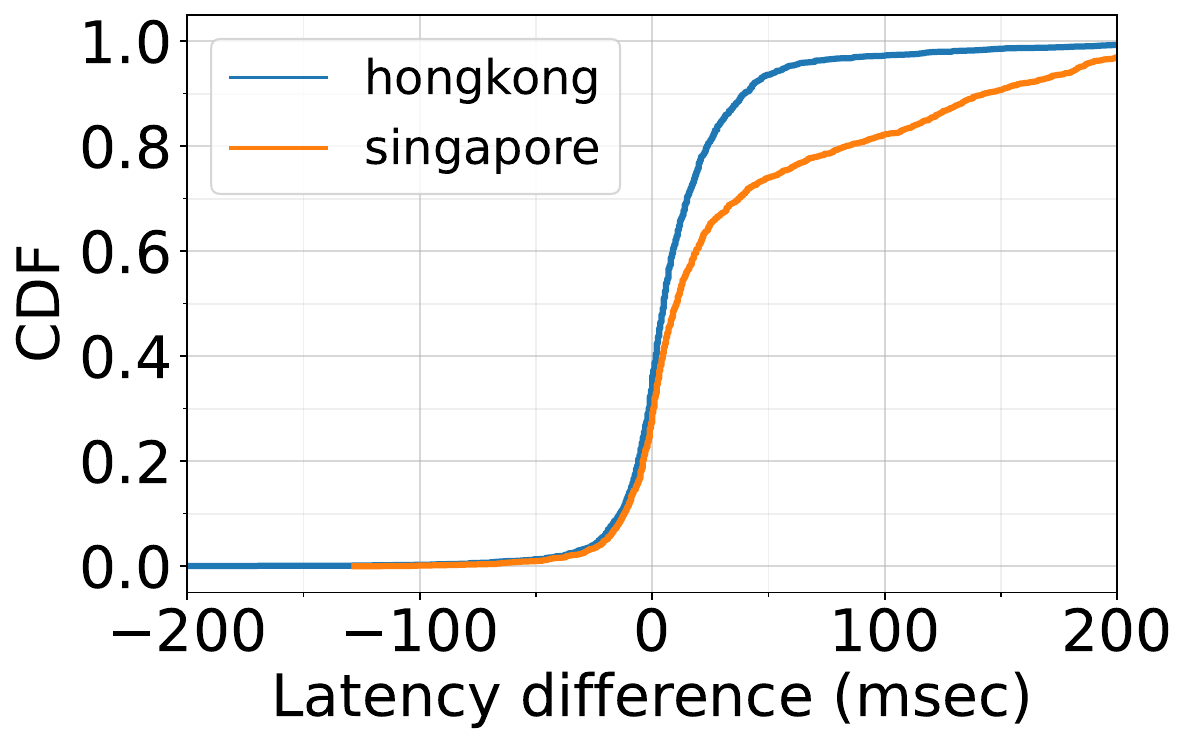}
\label{fig:measure:overall:con3}
}
\hspace{-0.3cm}
\subfigure[Australia + Africa]
{
\includegraphics[width = 0.24\textwidth]{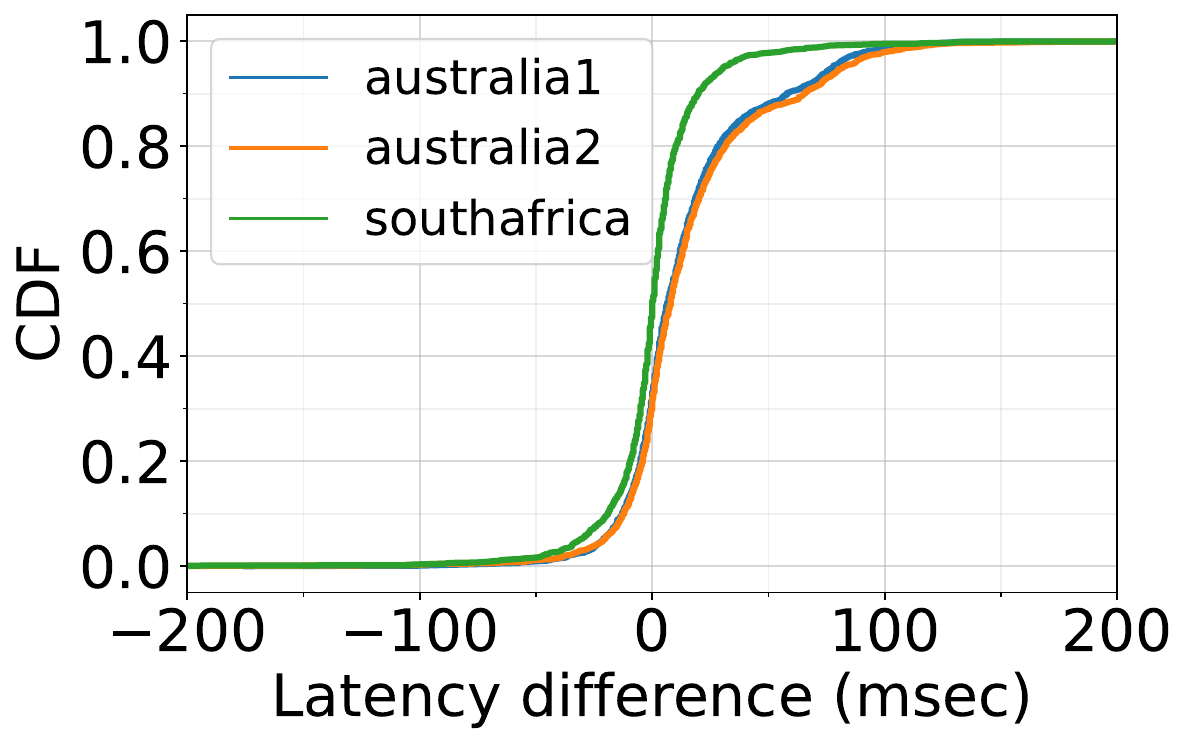}
\label{fig:measure:overall:con4}
}
\caption{Comparing latencies for WAN and Internet for 21 \azure DCs in 5 different continents. Negative difference indicates Internet is better. We label latency = RTT. The legends denote the DC locations.}
\protect\label{fig:measure:overall}
\end{figure*}

\textbf{Latency analysis}: For each hour, we calculate the median latencies between each client country and MP DC pair over Internet and WAN (we also consider finer granularity like ASN instead of country at the end of this section). We take the difference (Internet minus WAN) for these hourly median values from each client country to each MP DC. Fig.\ref{fig:measure:overall} plots the CDFs of differences for DCs in $5$ continents across all client countries for $7$ days in June 2024. \rohan{These results cover source and destination pairs comprehensively around the globe.} The key observations are:

(1) \db{In $33.73$\% cases,} Internet is strictly better than WAN. 

(2) 
In $23.98$\% cases, Internet is worse than WAN \db{by only up to $10$~msec.} 

(3) 
In $19.61$\% cases, Internet is worse than WAN \db{with a latency inflation between $10$ and $25$~msec.}

(4) For the remaining $22.68$\% cases, WAN latency is better than Internet by more than $25$~msec.

\db{As detailed in \S\ref{sec:overview:e2e}, user experience does not degrade substantially with a small increase in latency -- we set thresholds to $10$~msec and $25$~msec accordingly.}

\begin{figure*}[t]
\centering
\vspace{-0.2in}
\includegraphics[width=0.85\textwidth, page=1]{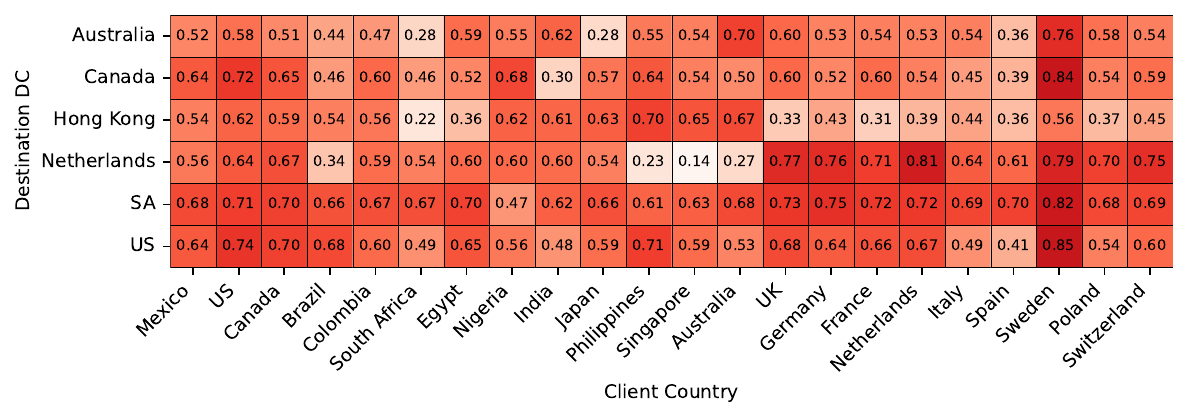}
\caption{Fraction (F) of times Internet provides better or comparable (within $10$~msec) latency compared to WAN. SA denotes South Africa and US denotes the United States. Darker shade means higher F.}
\label{fig:measure:latencyheatmap}
\end{figure*}

\textbf{Zooming in further:} Next, we deep dive into understanding the Internet routing low-latency opportunity for different geographic regions by quantifying the fraction ($F$) of times (when considering hourly median values for 1 week) Internet paths offer latencies lower than or comparable ($\leq${}$10$~ms inflation) to WAN paths from different client countries to destination DCs. Fig.\ref{fig:measure:latencyheatmap} plots the heatmap for paths between $22$ countries (spanning $5$ different continents; top $20$ by call volume and $2$ from Africa) and $6$ DCs from $5$ continents (Orange triangles in Fig.\ref{fig:measure:locations}). The key observations are:

(1) Internet paths often offer lower or comparable latencies in the North America (NA) -- Europe corridor. 
$F=41$-$85\%$ for Europe to NA and $64$-$74\%$ for NA to Europe paths. 

(2) Europe is well connected (low latency) to European and South Africa DCs over the Internet.
     
(3) Internet routing between Europe and the DC in Hong Kong performs poorly ($F=31$-$56\%$).

\textbf{Why is Internet better:} Internet performs well due to the well-provisioned trans-Atlantic fiber connectivity\cite{cables:web} offering similar latency choices to the Internet and WAN. Additionally, Internet provides better performance in some cases due to richer availability of peering points\cite{wan:infocom20}.

\textbf{Stability:}  We repeat the above experiment with data collected for a week but $6$ months in the past (Jan'$24$). We observe that, in $6$ months, the Internet has become slightly better for the NA - Europe corridor, while the broad trends hold true. The figure is available in \S\ref{sec:app:stability} (Fig.\ref{fig:app:1}).

\textbf{Long-term trends:} For both the Internet and WAN paths between the $20$ countries (top; by call volume) and all DCs, we measured the weekly median latencies for the weeks separated by 12 months. We found that in $80${}$+\%$ cases latencies have improved for both types of paths. The Internet paths see slightly greater improvements. More details are in \S\ref{sec:app:trend}.

\begin{wrapfigure}{l}{0.4\textwidth}
\centering
\includegraphics[width=0.4\textwidth, page=1]{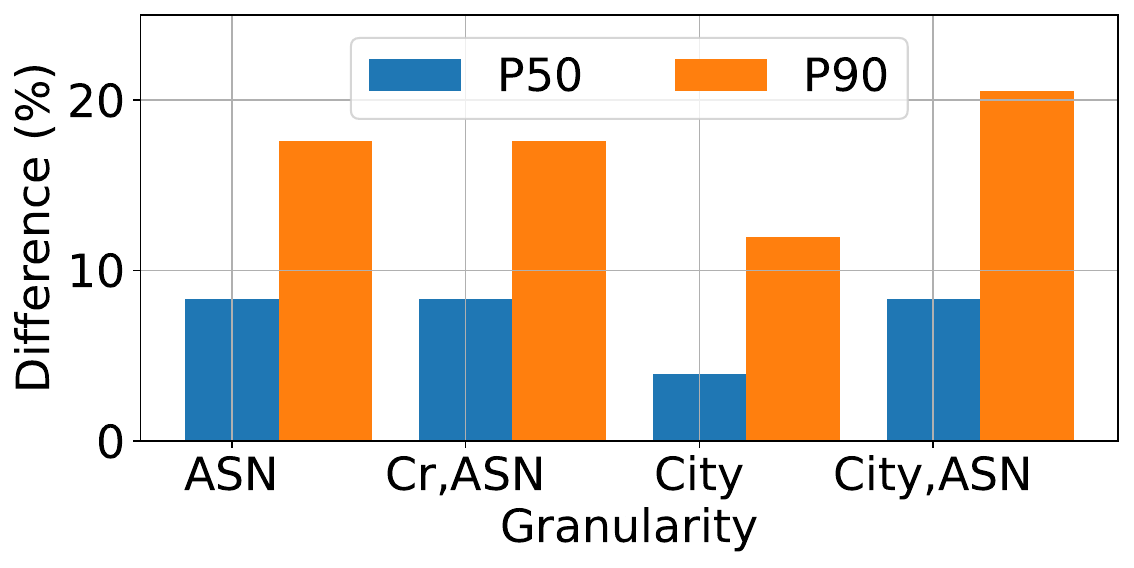}
\caption{Difference in $F$ between different granularities and granularity = country. Cr indicates country.}
\vspace{-0.1in}
\label{fig:measure:clarity}
\end{wrapfigure}

\textbf{Clustering using city and ASN (instead of country):} \db{Fig.\ref{fig:measure:latencyheatmap} shows the fraction ($F$) when we cluster the measurements at the country granularity. We do similar analyses at the clustering granularity of cities, ASNs, and city + ASNs. For ASN (similarly other) clustering, we take the weighted difference compared to clustering using country. The weights are the fraction of measurements for each ASN in the country (more details in \S\ref{sec:app:gran}). The results (fraction $F$) do not change significantly compared to the country-level granularity (difference bound to $8\%$ at $P50$) as shown in Fig.\ref{fig:measure:clarity}.}  

In summary, Internet \db{latency} is better or comparable to WAN in parts of world. Also, the measurements are stable and 
\db{the insights hold true} even at the granularity of countries \db{thus making us cautiously optimistic} about moving a fraction of the \teams traffic to the Internet.

\section{\tilda: Moving calls to Internet}
\label{sec:intmigration}
 
Based on the latency measurements in the previous section, we select the client countries and MP DCs in \textit{Europe} as candidates to move some of the traffic to the Internet. In this section, we detail \tilda\xspace -- our system that carefully moves a fraction of the traffic from WAN to the Internet without hampering user experience. \tilda has been in production for the last $12$ months. 

\subsection{Determining fraction of traffic to move}
\label{sec:intmigration:fraction}

The key challenge when moving \teams traffic to the Internet is that we do not know the capacity of such paths in advance. Na\"ively moving all \teams traffic to the Internet could cause network congestion and, hence, a poor user experience. We prioritize \textit{safety} over optimality -- we stop moving traffic to the Internet, even if there is no performance degradation. Our design for carefully moving traffic involves the following key elements. 

($1$) \textbf{Granularity of the movement:} \tilda  moves traffic to the Internet at various levels of granularity, from a small number of users, metro, ASN to the country level. We cautiously start with small communities of \teams users and move entire country if the performance is acceptable.  

($2$) \textbf{A|B testing:} Regardless of the granularity of the movement to the Internet, we use an Experimentation and Configuration System or ECS that conducts A|B experiments on a percentage of the user population and generates scorecards to analyze and control the traffic shift.

($3$) \textbf{Variable traffic allocation:} For each combination of client country and MP DC, we typically increment $1$-$3\%$ (based on domain knowledge) of the traffic, at a time, to the Internet. After each move, we monitor the performance metrics for a few days for stability and make quick reactions (detailed next) after observing negative effects. Otherwise, we repeat the process, moving more traffic for the pair to the Internet. We currently stop at 20\% based on operational expertise. Each MP DC is connected to Internet via multiple transit providers. When calculating \% movement: (a) we consider the minimum capacity available on \azure links peering with the transit providers. (b) we assign different priorities to client countries (based on importance) and split available (minimum) capacity across client countries depending on their priorities. 

($4$) \textbf{Quick reaction to poor performance:} We decrease/stop moving traffic to the Internet when performance criteria for some of the network/application metrics (including network latency, loss or jitter, or application MOS) are not met. We do so instantaneously. We continually monitor the network network metrics as calls progress. We collect MOS  (Mean Opinion Score; user feedback)  at the end of a subset of calls.

While determining points of congestion on the Internet is a challenging task (and a subject for further study), we have a few knobs to react to poor performance quickly: (a) If a large fraction of calls over the Internet in a client country -- MP DC mapping shows moderate performance degradation across some of the metrics (e.g., P50 packet loss $\leq$ 1\%, latency inflation $\leq$ 10\%), we decrement traffic on Internet for that client country -- MP DC pair. (b) \textit{emergency breaks:} If there is a severe performance degradation (e.g., P50 packet loss $\geq$ 1\%, rare), we reroute traffic over the WAN., (c) If only a few users  are facing problems, we move them selectively to WAN as detailed in \S\ref{sec:design:realtime}. 
(d) If high degree of unavailability is observed from one MP DC to a transit ASN, network automatically mitigates by failing over to alternative transit peer (using BGP). If mitigation is not complete due to some constraints, we will detect the anomaly and failover to the WAN.

($5$) \textbf{Traffic to be moved:} With \tilda, we randomly select the call participants to be assigned the Internet paths, constrained by the fraction chosen for the client country and the target MP DC. 

Overall, the \tilda system involves a combination of A|B testing, performance analysis, rule-based control, and the ability to adapt dynamically based on network conditions.

\subsection{\tilda production findings}
\label{sec:intmigration:experiences}

We now detail our production experiences while moving large-scale traffic from WAN to Internet.

\begin{figure*}
\centering
    \begin{minipage}{.45\textwidth}
        \centering
        \includegraphics[width=0.95\textwidth, page=1]{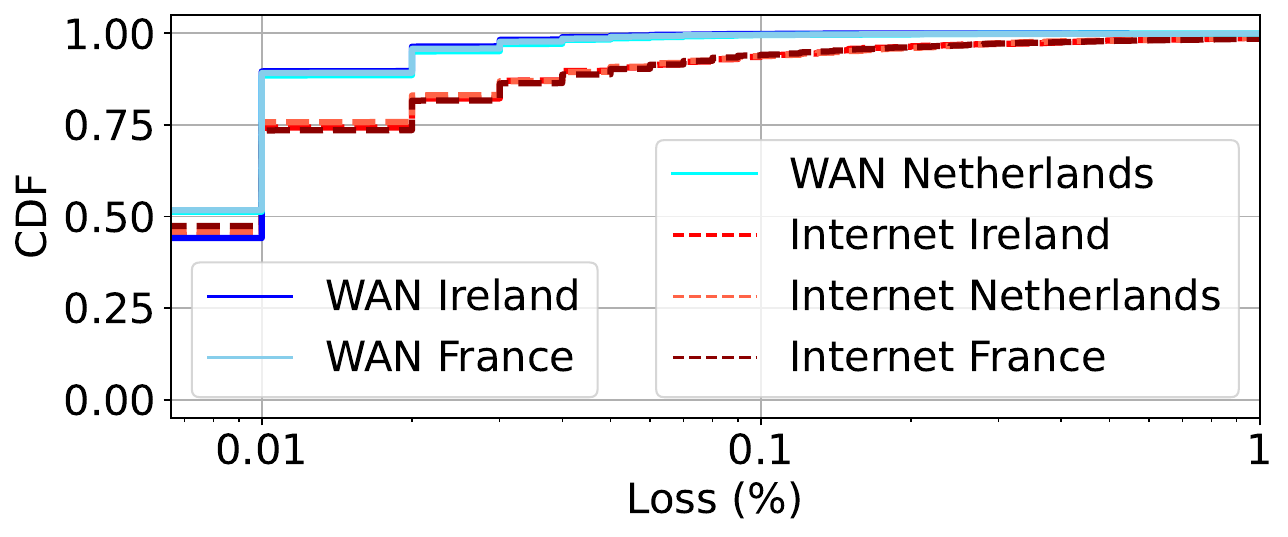}
        \caption{Packet loss for Internet and WAN.}
        \label{fig:measure:loss}
    \end{minipage}%
    \hspace{0.1cm}
    \begin{minipage}{0.45\textwidth}
        \centering
        \includegraphics[width=0.95\textwidth, page=1]{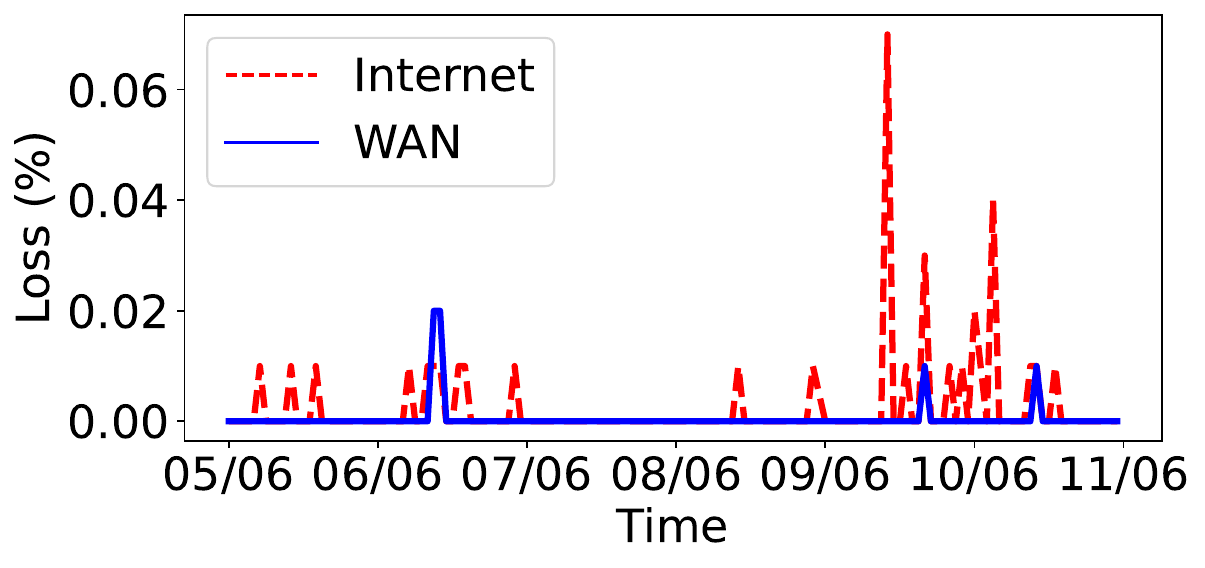}
        \caption{Packet loss between France and DC in the Netherlands.}
        \label{fig:measure:losstime}
    \end{minipage}%
\end{figure*}

\textbf{(1) The Internet has worse loss (in general):}
We pick $3$ DCs in Europe (Ireland, Netherlands, and France) for which a fraction of the \teams traffic is moved to the Internet, and log the average loss reported by RTP~\cite{rtp:tech2003} (using missing sequence numbers) for each call participant in Europe. For each client country-MP DC pair, we find the hourly median loss for $7$ days between $5^{th}$ and $11^{th}$ June'$24$. Fig.\ref{fig:measure:loss} shows the CDFs of hourly median loss for traffic between the $3$ DCs and all client countries in Europe over WAN and the Internet. While it is evident that loss rates are low ($\leq0.01\%$) for a large fraction of both the Internet (average of $44.9\%$ across DCs) and WAN paths ($49.2\%$, likewise), the Internet paths have higher loss rates at the tail. For $\sim${}$10\%$ cases, the Internet paths could experience at least $0.1\%$ loss, while such loss over the WAN is almost non-existent.

\textbf{(2) The Internet has more loss spikes:} Fig.\ref{fig:measure:losstime} shows the time series for hourly median loss rates between the Netherlands DC and clients in France. Internet paths have higher (up to $3\times$) and more frequent loss spikes than WAN, with the peak loss rates for the latter limited to only $0.02\%$. While \teams can mitigate loss to a some degree by using application layer redundancy mechanisms, one should be cautious in moving traffic to the Internet so as not to risk performance degradation due to inflated losses. The trends are similar for other client country -- MP DC pairs as shown in Fig.\ref{fig:app:loss} (\S\ref{sec:app:loss}). We measure the number of 30 minute timeslots over 7 days observing at least 0.1\% (and 1\%) loss on Internet and WAN for all client countries -- MP DC pairs in Europe. Fig.\ref{fig:app:loss} in \S\ref{sec:app:loss} shows that Internet has more frequent loss.

\begin{wrapfigure}{l}{0.45\textwidth}
\centering
\vspace{-0.1in}
\includegraphics[width=0.45\textwidth, page=1]{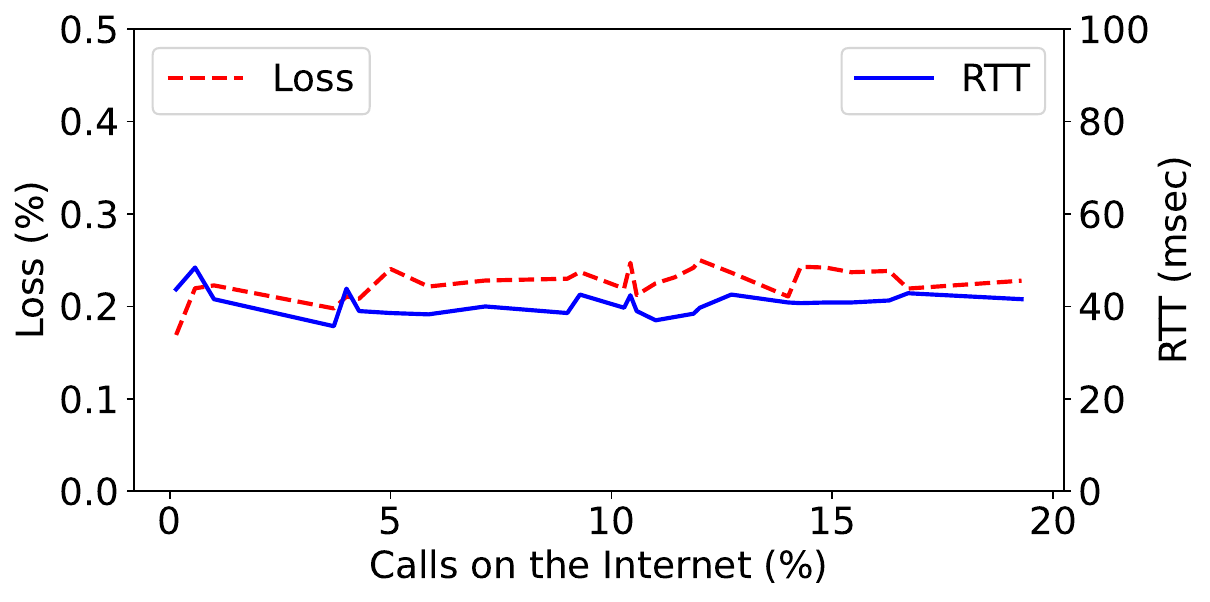}
\caption{Loss, RTT vs. fraction of traffic on Internet between the UK users and Netherlands DC.}
\label{fig:measure:loss_rtt}
\vspace{-0.2in}
\end{wrapfigure}

\textbf{(3) Internet has higher jitter:} We have observed that the Internet has slightly worse jitter than WAN up to $10\%$. We have observed mean jitter of $3.4$~msec and $3.52$~msec for WAN and Internet paths in North America region. Note that, \teams uses jitter buffers\cite{bitag:latency2022} and this additional jitter on Internet does not affect performance.

\textbf{(4) The Internet is reasonably elastic:} Fig.\ref{fig:measure:loss_rtt} shows the loss and RTT vs. fraction of traffic moved to the Internet between clients in the UK and the MP DC in the Netherlands. Note that, even when $20\%$ of \teams traffic is moved to the Internet, neither packet loss rate nor latency shows any systematic inflation. We repeat the same experiment for all client country -- MP DC pairs in Europe. The median change for latency and loss are 3 msec and 0.06\% (Fig.\ref{fig:app:elasticity} in \S\ref{sec:app:elasticity}). These results show that Internet is reasonably elastic for a large number of client country -- MP DC pairs, demonstrating \tilda not adding significantly to the Internet congestion. Note that at fractions higher than $20\%$ (not tried in production), there is a chance that we congest the Internet paths thus inflating loss and latency.

\textbf{(5) When the Internet is not an option:} We have observed Internet paths for some client countries (e.g., Germany, Austria) with high (and unacceptable) loss even when a small amount of traffic was moved. In such cases, we do not use the Internet at all. At the same time, for some MP DCs, we have observed high transient loss from a few client countries, which was affecting user experience. Thus, we stopped moving traffic to Internet for those MP DCs, and relied on the WAN.

\textbf{(6) \db{Congestion \textit{likely} at the transit ISPs}:} We detected variations in performance across transit ISPs between DCs and users (Fig.\ref{fig:back:hot}). We observed  higher packet loss along end-to-end paths between an MP DC and multiple ISPs simultaneously. Such one-to-many loss patterns, with no corresponding loss inflation observed at the DC or the WAN, hint at congestion at the transit ISPs. We had to react to performance degradation by steering traffic to alternate transit providers.

\textbf{(7) Internet as a fall-back option:} Some of our WAN cables observed a fiber-cut and did not recover for months. As a result, our WAN capacity to Africa dropped to just a few hundreds of Gbps. 
As Internet was providing comparable performance, we moved some of our \teams's traffic to Internet with \tilda to make more capacity available on WAN for other services. In a different case, we used Internet to onboard new customers while WAN capacity was getting built.

\section{Joint assignment in \name}
\label{sec:overview}

As quantified in the previous section, we can move a subset of the \teams traffic to the Internet with fairly good performance. At the same time, the cost of using a WAN link depends on the \textit{peak} usage\cite{switchboard:sigcomm23, pretium:sigcomm16}. Therefore, the total \teams traffic needs to be split across WAN and Internet to reduce peaks on WAN (and operating costs) while not impacting performance.

\tilda assumed that the MP DC assigned to the call is fixed (assigned outside \tilda). \tilda calculates fraction of traffic on Internet paths, and simply assigns participants to Internet paths \textit{randomly} based on the fractions. However, MP DC assignments done outside \tilda may likely be unaware of the Internet offload opportunities and may make sub-optimal assignments (\S\ref{sec:overview:joint}). We build \name to make efficient use of the Internet paths by jointly assigning the MP DC and the routing option to individual calls. We next discuss the two key ideas in \name.

\subsection{Joint MP placement and routing} 
\label{sec:overview:joint}

\begin{SCfigure}[1][t]
\centering
\includegraphics[width=0.45\textwidth, page=2]{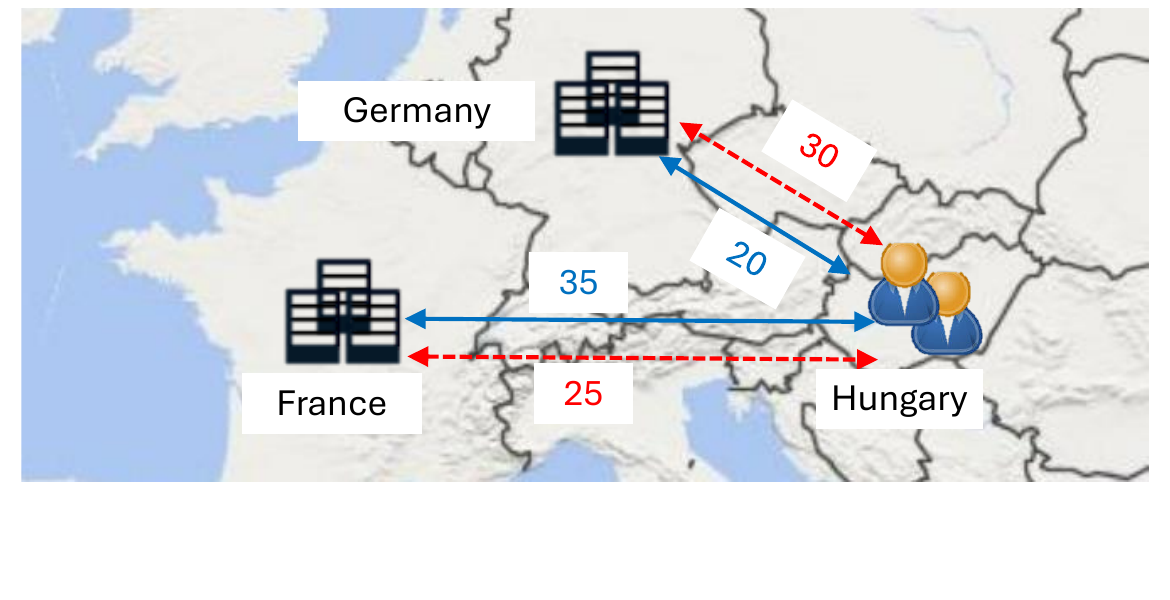}
\vspace{-0.3in}
\caption{Benefits of joint optimization. There is a call with users in Hungary with potential MP DCs in France and Germany. Blue and red arrows indicate WAN and the Internet paths; numbers indicate latency (msec).}
\label{fig:key:joint}
\end{SCfigure}

A strawman's approach could be to use Switchboard~\cite{switchboard:sigcomm23} (based on locality) to first calculate the MP DCs for calls and then move traffic to the Internet if such paths have capacity. However, such a scheme may result in sub-optimal assignment as shown in Fig.\ref{fig:key:joint}. Imagine a call with $2$ users in Hungary and potential MP DCs in France and Germany. The latencies for Internet and WAN paths are shown in the figure. Switchboard, synergistic to locality (nearest DC) may select the DC in Germany by looking at the WAN latencies. Once this DC is picked for MP allocation, we then move the call to the Internet if there is capacity. But this scheme will result in a latency of $30$~msec. However, assigning the call to the DC in France with the Internet routing option could have provided lower latency ($25$~msec) while reducing WAN overhead. Alternately, we may formulate Switchboard to use the Internet paths when calculating MP DCs, and then move calls to use WAN paths, which, similarly, is also sub-optimal. Thus, instead of assigning MP DCs and Internet routing separately, in \name we \textit{jointly make such assignments} toward minimizing WAN peaks. \rohan{In doing so, we continue to provide same bandwidth to the users \db{at lower or comparable latency}.}

\begin{figure*}
\centering
    \begin{minipage}{.48\textwidth}
        \centering
        \includegraphics[width=0.95\textwidth, page=1]{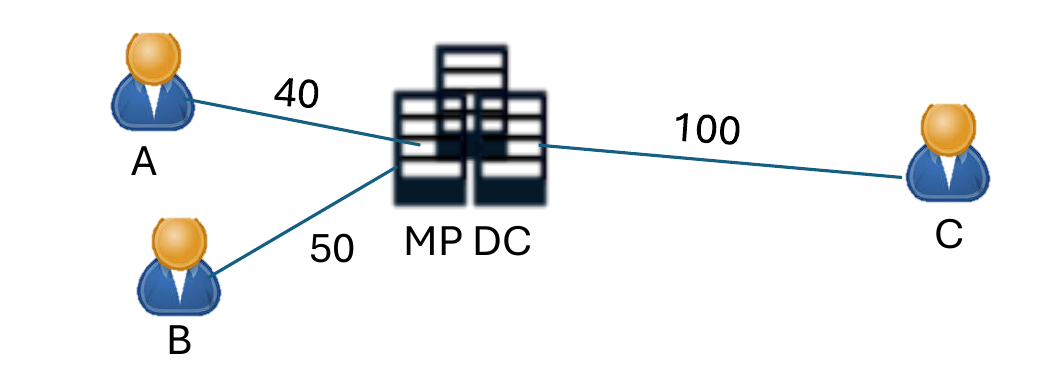}
        \vspace{0.15in}
        \caption{E$2$E latency is important in MP DC and routing assignment.}
        \label{fig:key:E$2$E}
    \end{minipage}%
    \hspace{0.1cm}
    \begin{minipage}{0.46\textwidth}
        \centering
        \includegraphics[width=0.95\textwidth, page=1]{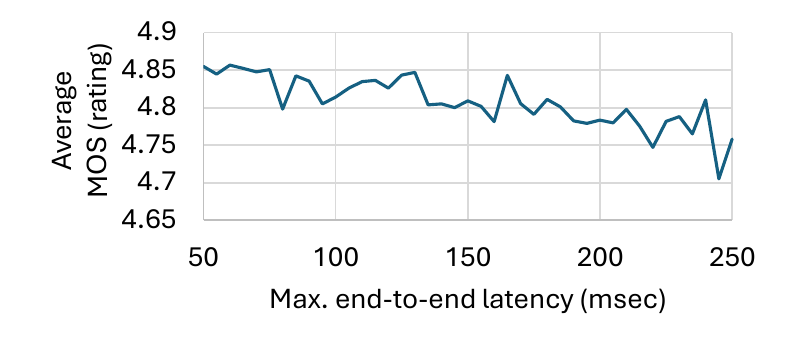}
        \caption{Impact of max. E$2$E latency on user experience.}
        \label{fig:measure:ux}
    \end{minipage}%
\end{figure*}

\subsection{Optimizing for end-to-end latency} 
\label{sec:overview:e2e}

The experience of any two users engaged in a conversation depends on the end-to-end (E$2$E) latency between them (e.g., $50 + 100 = 150$~msec for users B and C in Fig.\ref{fig:key:E$2$E}). Under the assumption that any two users can engage in a conversation during a call, the \textit{maximum} E$2$E latency across all participants determines the participant engagement and, hence, the user experience in such calls.

To understand the impact of max. E$2$E latency on user experience for \teams, we leverage the MOS (Mean Opinion Score; user feedback) as a measure of the Quality of Experience seen by the user (In contrast, \cite{mos:hotnets23, switchboard:sigcomm23} do not show impact of E$2$E latency on MOS).
 
\teams telemetry also collects the latency between MP and the participant. Based on the latency reported for each participant, we calculate the max. E$2$E latency and group all assigned MOS for $5$~msec  buckets of max. E$2$E latency. Fig.\ref{fig:measure:ux} shows the average MOS for increasing max. E$2$E latency. We select the range of $50$-$250$~msec as we have significant measurements (at least $1$,$000$ points for each $5$~msec bucket). \textit{Note that MOS is collected at the end of a subset of calls and is heavily sampled.} It can be seen that: (a) For max. E$2$E latency under $75$~msec, the impact on MOS is minimal indicating that users are tolerant of E$2$E latency up to a certain extent. (b) The user experience degrades (mostly linearly) with an increase in the max. E$2$E latency. Thus, service operators are keen to keep this E$2$E latency low. We take max. E$2$E latency into consideration when assigning the MP DCs and routing options to individual calls to bring users (virtually) closer and improve user experience.

\section{\name design}
\label{sec:design}

\textbf{Inputs:} \name needs to assign the MP DC and routing option when the call starts, \ie when the first user of the call joins\footnote{We cannot do the assignments when the second or subsequent participants join as there are calls with single participant that use other features of \teams (e.g., video recording, transcript) that requires MP assigned.} \db{based on the location (country) of the first joiner. If needed, The MP DC and routing option could be changed later (\S\ref{sec:design:migration}).}

The other inputs used by \name are: ($a$) the numbers of available MPs in individual DCs (fixed; discussed in \S\ref{sec:back:mp}), ($b$) rich history (participant's country, media types, time of the call; anonymized) of calls to predict the peaks and make assignments accordingly, ($c$) \db{Internet path capacities for each client country - MP DC pair as recorded by \tilda}, and ($d$) WAN topology and Internet peering points (\teams is a first party conferencing service with access to such details).

\textbf{Call config:} We want to assign the MP DC and routing option for each individual call. To do such assignments at scale, we borrow the notion of \emph{call configuration} (call config, for short) from Switchboard~\cite{switchboard:sigcomm23}, which captures the resource requirements of calls through the number of participants and media types of different calls. A call config comprises ($1$) the location (country) of the participants, ($2$) participant count from each country, and ($3$) media type (audio, video, or screen-share). A call can have any of the above media streams, but we assign call config using the most resource-hungry media type (audio $<$ screen-share $<$ video). An example call config is \emph{((France-$2$, UK-$1$), Audio)} which represents all audio-only calls with $2$ participants from France and $1$ from the UK. All calls with the same call config are fungible -- they have largely the same resource requirement. The number of call configs is orders of magnitude fewer than the number of calls that helps scale the LP (details in \S\ref{sec:lp}).

\subsection{\name building blocks}
\label{sec:design:bblocks}

\begin{wrapfigure}{r}{0.45\textwidth}
\centering
\vspace{-0.15in}
\includegraphics[width=0.45\textwidth, page=1]{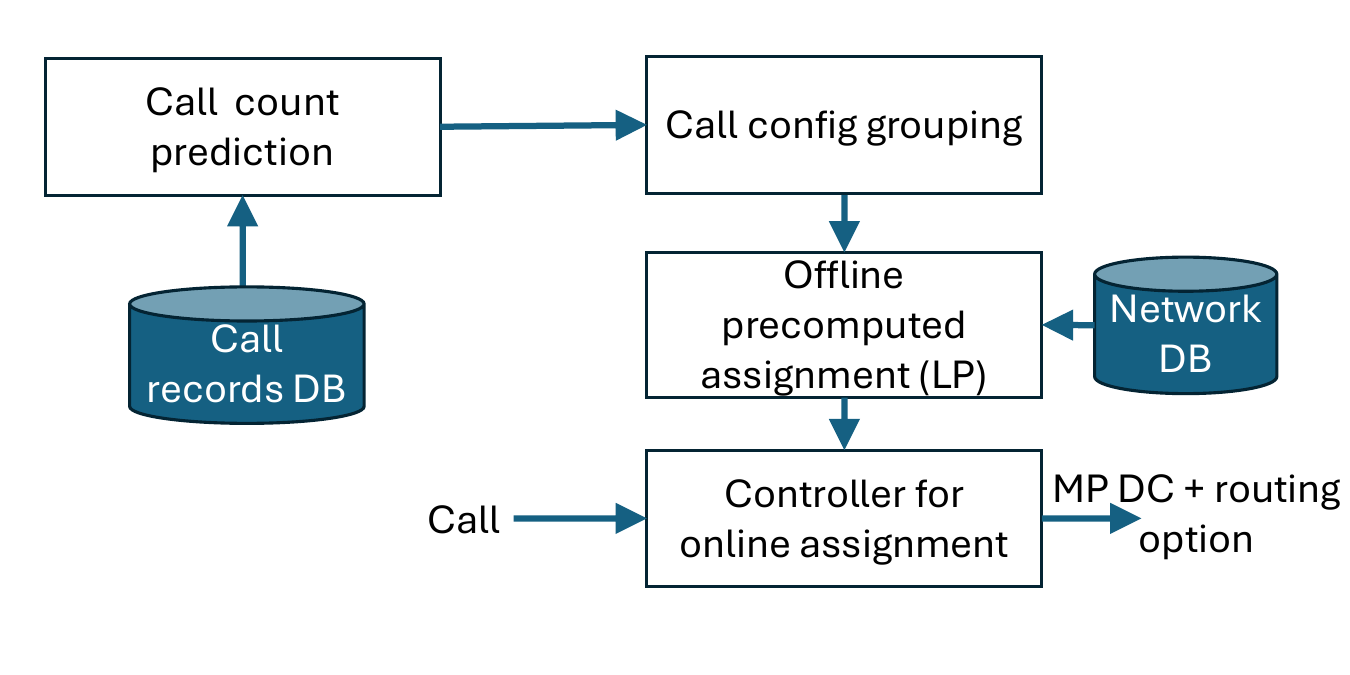}
\caption{Building blocks in \name}
\label{fig:design:overview}
\vspace{-0.2in}
\end{wrapfigure}

Fig.\ref{fig:design:overview} shows the building blocks in \name. In a nutshell, \name pre-computes an offline assignment plan, based on the expected (predicted) call demand. It uses such a plan to assign calls at run-time. This way, using prediction, the offline plan can do the assignment to minimize  peaks in the network links. \name has five modules:

($1$) \textbf{Call records database}:
    \teams records and stores some data (anonymized) for each participant of the call
    including the start time, media type, time of the call, MP DC country, and the latency experienced by the user (client-to-MP). \name uses these call records to forecast demands as well as to calculate latencies of call participants. 
    
($2$) \textbf{Call count prediction}:
\name assigns the MP DC and routing option for (reduced) call configs (explained next). To do so, for each call config, \name uses Holt-Winters exponential smoothing\cite{hw:web} to forecast the number of calls for the next $1$~day at the granularity of $30$~min timeslots. \name predicts for the top $3$,$000$ call configs covering $90${}$+\%$ of all calls (to finish prediction quickly). \db{For each $24$~hour prediction with high accuracy (\S\ref{sec:evalcontroller:prediction})}, we use $4$~weeks of training data. 

($3$) \textbf{Call config grouping:} 
The difference in MP DCs assigned to call config vs. assignment using the country of the first joiner may lead to call migration from one MP DC to another. To reduce such migrations, we transform all calls to their \textit{reduced call config} and group all call configs falling into the same reduced call config. We discuss this next in \S\ref{sec:design:migration}.
    
($4$) \textbf{Offline precomputed plan:} Using the forecasts, reduced call configs, and network database consisting of WAN topology and Internet path capacity, this module assigns the pairs of MP DC and routing option for each reduced call config for each 30 minute timeslots for next 24 hours. We formulate it as a Linear Program (LP) to minimize the sum of peaks on individual WAN links (\S\ref{sec:lp}). 

($5$) \textbf{Controller for online assignment:} Given the pre-computed assignment from ($4$) above, this module assigns the MP DC and routing option to each incoming call when the first participant of a call joins. We use a combination of offline pre-computed plan and the country of the first participant. Moreover, we might need to migrate the call to another DC if the initial assignment is not according to the pre-computed plan. The controller is very fast when assigning the MP DC and routing option and does not cause any performance degradation.

\subsection{Reducing call migrations}
\label{sec:design:migration}
 
\tilda can make MP DC assignment using LP in Switchboard\cite{switchboard:sigcomm23}.
Imagine a call where the first joiner is from Germany, and the LP has a pre-computed plan for both (Germany-$2$, Audio) and (Germany-$3$, Audio) call configs. Because the LP makes decisions for individual call configs, such assignments might end up assigning different MP DCs for these two call configs. E.g., (Germany-$2$, Audio) is assigned to Ireland and (Germany-$3$, Audio) is assigned to France. When the first user joins, we do not know the call config (we only know the country of the first joiner). Yet, we need to assign the MP DC, either Ireland or France. Let's say we assign the call to Ireland and as the call progresses, we realize that the true call config is rather (Germany-$3$, Audio) which ought to be assigned to France to adhere to the pre-computed plan. Thus, we need to \textit{migrate the call} from Ireland to France. Such migrations are undesired as they result in user-perceived glitches.

We minimize migrations in \name using the following mechanism: we first \textit{transform} call configs to factor out the \textit{scale} (number of participants from one country) from the \textit{distribution} (participants across countries in call config). In general, we transform the call config such that the number of participants from each country in the call config has a GCD of $1$. For intra-country calls, we transform the call configs to just have $1$ participant (e.g., (Germany-$2$, Audio) is changed to (Germany-$1$, Audio)). We keep the resource requirement the same. E.g., let's say there are $100$ calls with config (Germany-$2$, Audio). We transform it to $200$ calls with config (Germany-$1$, Audio). We call these new configs \textit{reduced call configs}. We then \textit{group} together all calls based on reduced call configs (e.g., (Germany-$2$, Audio) and (Germany-$3$, Audio) are grouped together using the reduced call config (Germany-$1$, Audio)). We do not group call configs across media types as they have different network and compute requirements. The LP makes decisions at the granularity of this reduced call config. This way, we significantly reduce call migrations arising due to differences in assignments for call configs from the same country. However, as detailed in \S\ref{sec:lp} and \S\ref{sec:design:realtime}, we do not eliminate migrations due to differences in media types and for international calls.

\subsection{MP and routing assignment}
\label{sec:lp}

\begin{table*}[t]
\begin{center}
\caption{Notations used in the LP.}
\small
\begin{tabular}{|c|c|}
\hline
\textbf{Notation} & \textbf{Definition}\\
\hline
$T$, $M$, $C$ & Set of timeslots, set of MP DCs, set of reduced call configs\\
\hline
$Cap_{t,m}$ & Compute capacity of the MP DC $m$ for the timeslot $t$ in terms of number of cores\\
\hline
$I, W$ & Set of Internet paths and set of WAN links\\
\hline
$P$ & Set of all Internet and WAN paths. Each MP DC has one Internet and WAN path each\\
\hline
$N_{t,c}, N$ & ($N_{t,c}$) Number of calls for the call config $c$ for the timeslot $t$, $N$: total number of calls\\
\hline
$InternetCap_{t,p}$ & Capacity (in Gbps) of the Internet path $p$ in the timeslot $t$ \\
\hline
$E$ & Bound on the average of max. end-to-end latency across reduced call configs\\
\hline
\hline
(output) $X_{t,c,m,p}$ & Number of calls assigned to $c$-th call config to the MP DC $m$ and path $p$ for timeslot $t$\\
\hline
(output) $y_{l}$ & peak bandwidth used on the WAN link $l$\\
\hline
\end{tabular}
\label{tab:design:lp:notations} 
\end{center}
\end{table*}


\begin{figure*}[t]
  \centering
{\small
\fbox{
  \begin{minipage}{0.965\textwidth}
    \textbf{LP Variable:} $X_{t,c,m,p}$ \hspace{0.5in}
    \textbf{Objective:} Minimize $\textstyle\sum_{l \in W} y_{l}$\\
    \textbf{Constraints:}\\
    $\color{blue}C_1$: $\forall t \in T, c \in C, \textstyle\sum_{m \in M, p \in P} \hspace{0.3em} X_{t,c,m,p} = N_{t,c}$\\
    $\color{blue}C_2$: $\forall t \in T, m \in M$, $\textstyle\sum_{c \in C, p \in P} \hspace{0.3em} X_{t,c,m,p} \cdot computeUsed(c) \leq Cap_{t,m}$\\
    $\color{blue}C_3$: $\forall t \in T, p \in I$, $\textstyle\sum_{c \in C, m \in M} \hspace{0.3em} X_{t,c,m,p} \cdot networkUsed(c,m,p)  \leq InternetCap_{t,p}$ \\    
    $\color{blue}C_4$: $\frac{1}{N} \cdot \textstyle\sum_{t \in T, c \in C, m \in M, p \in P} \hspace{0.3em} X_{t,c,m,p} \cdot E2Elatency(c,m,p) \leq E$ \\
    $\color{blue}C_5$: $\forall t \in T, l \in W$, $y_{l} \geq \textstyle\sum_{c \in C, m \in M, p \in P} \hspace{0.3em} X_{t,c,m,p} \cdot networkUsed(c,m,p) \cdot isLinkUsed(c,m,p,l)$ 
  \end{minipage}
}
}
\caption{LP formulation for joint MP DC and routing option assignment. Notations are in Table \ref{tab:design:lp:notations}.}
\protect\label{fig:algo:lp}
\end{figure*}

We now detail our LP  to jointly calculate the MP DC and routing option for the reduced call configs. The LP is shown in Fig.\ref{fig:algo:lp} with notations described in Table \ref{tab:design:lp:notations}.

\textbf{LP objective:} The objective of the LP is to minimize the sum of peaks on the WAN links (leveraging Internet paths and MP DC selection). The peaks are calculated for $24$ hours period. This directly reduces  network costs for \teams. 

\textbf{Frequency:} We run the LP every $30$~min (with fresh estimates) that calculates the assignments for the next $24$~hours (to make assignments aware of the daily peak) in $30$~min time-slots. This approach: (a) effectively reduces the WAN traffic peak while keeping traffic on the WAN during off-peak hours 
and (b) by running every $30$~min, it adapts the assignments to fresh information about the fraction of traffic on Internet calculated by \tilda. The LP does not have details about the run-time conditions (loss and latency) for the participants on the individual calls apriori.
\name adapts to these run-time conditions as detailed in \S\ref{sec:design:realtime}.

\textbf{LP variable:} The LP variable is $X_{t,c,m,p}$ that indicates the number of calls for reduced call config $c$ in timeslot $t$ assigned to the MP DC $m$ over the path $p$. We have two options for $p$ for each MP DC -- the Internet vs. WAN. When assigned to WAN/Internet, the underlying routing algorithm (outside the scope of \name) decides the path to the destination.

\textbf{Constraints:} We have five constraints as follows:

($\color{blue}C_1$) Total number of calls: For each reduced call config $c$ and timeslot $t$, we assign MP DC(s) and paths (possibly multiple combinations) to all calls of that config.

($\color{blue}C_2$) Compute capacity: For each MP DC, we assign the calls such that the total compute capacity of the MP DC is not exceeded in each timeslot. The $computeUsed()$ function (Fig.\ref{fig:algo:lp}) returns the compute required for $c$ based on its media type and the number of participants. $Cap_{t,m}$ denotes the compute available in the MP DC $m$ in timeslot $t$.

($\color{blue}C_3$) Internet capacity: We assign $m$ and $p$ such that the Internet capacity is not exceeded for any of the Internet paths and timeslots. We estimate the Internet path usage by multiplying the fraction of calls to be moved to the Internet, the total number of participants, and the average network usage per participant. 
The $networkUsed()$ function estimates the bandwidth consumed by $c, m, p$ using its media type and the number of participants.

($\color{blue}C_4$) End-to-end (E$2$E) latency: We do the assignments so that \textit{average} of max. E$2$E latency across call configs is bounded. $E2Elatency()$ function returns the max. E2E latency given a combination of $c$, $m$, and $p$. We tried putting a bound on maximum of max. E$2$E latency for each call config, but we observed that such a bound is stretched due to a handful of configs and is not useful for a majority of the configs. Thus, we choose to use a bound on average of max. E$2$E latency across call configs.

($\color{blue}C_5$) Denoting $y_{l}$: This constraint ensures that $y_{l}$ is set to peak link utilization on WAN link $l$ across all timeslots. $isLinkUsed()$ denotes whether $l$ is used.

\textbf{What did not work:} Majority of the calls today are \textit{intra-country} for \teams. We need to assign all calls from the same country to the same MP DC to eliminate call migrations between DCs. We formulated it as an ILP (Integer Linear Program) and the intra-country migrations did fall to zero. However, the network savings also substantially diminished as calls could not be assigned to multiple DCs to save network bandwidth. 
Consequently, we aim to reduce the number of migrations in \name instead of eliminating them altogether.

\textbf{Discussion:} As shown in Fig.\ref{fig:measure:loss_rtt} (and Fig.\ref{fig:app:elasticity} in \S\ref{sec:app:elasticity}), packet loss on the Internet does not increase significantly as we increase \teams traffic on the Internet \rohan{up to a certain extent}. Also, real packet loss is known only when calls progress. Thus, LP does not consider packet loss during assignments. We detail more on this in \S\ref{sec:design:realtime}. Secondly, the LP assigns \textit{single} routing option (either WAN or Internet) for all participants of the same call. Without this condition, LP size increased substantially and could not finish in timely manner. Lastly, we don't split  traffic from same participant across WAN and Internet links to avoid adverse effects at the receiver, especially for out-of-order packets and the jitter buffer. We leave such traffic splitting for future work.

\subsection{Real-time call assignment}
\label{sec:design:realtime}

\textbf{Initial assignment:} The precomputed plan above assumes full knowledge of the reduced call config. We need to assign the MP DC and routing option when the call starts, \ie when the first user of the call joins. However, we do not know the full call config when the call starts. We address this challenge as follows: recall that we transform each call config to a reduced version, and a  reduced call config can still have different assignments for different media types. For example, (Germany-$1$, Audio) could be assigned to Ireland, while (Germany-$1$, Video) is assigned to France. For a new call, we assume it as an intra-country call (such calls are in majority) and pick the \textit{most recently used} reduced call config based on the country of the first joiner. We then use all the counts for each assignment (MP DC and routing option) calculated by the LP for that reduced call config as weights and use weighted random to pick the assignment.

\textbf{Migration to different MP and routing option:} It may happen that even the reduced call config turns out to be incorrect as the call progresses. Consequently, the choice of the MP DC and routing option could also be incorrect. In such cases, we \textit{migrate} the call across MP DCs and routing options. To do so, we wait for $5$ minutes (configurable) into the call for the reduced call config to converge (e.g., initial call config = (Germany-$1$, Audio), while the converged call config = ((Germany-$1$, France-$1$), Video). \db{If the MP DC and routing option for reduced call config (after $5$ minutes) is different than the initial assignment, we migrate the call to the target assignment.} 

\textbf{Handling surge in calls:} We have rarely witnessed sudden high jump in calls. In such cases, MP servers are to be scaled accordingly. If we witness calls for which LP hasn't assigned capacity, we assign MP DC closest to the first joiner of a call that has enough capacity.

\textbf{Migration to a different route:} Recall that LP based assignment is offline -- it does not know the real-time conditions of the Internet paths. It may happen that the performance of the Internet path for a participant on a call is poor due to outages or transient congestion (e.g., frequent packet loss as shown in Fig.\ref{fig:measure:losstime}), and we want to react in seconds (cannot wait $30$~min for LP to address it -- would affect user experience). We monitor the packet loss and latency on the Internet path as the call progresses, and move the user to WAN  when the latency and packet loss are above acceptable thresholds: packet loss $\geq1\%$ and latency threshold is set depending on the physical distance. We observed the median number of users across $2$ months with loss on Internet $\geq1\%$ as $3.96\%$. In rare events when a large chunk of users experience poor performance, \tilda would take charge, adjusting the percentage of traffic on the Internet/WAN paths, and \name would simply abide. We do not move calls from WAN to Internet as to satisfy the Internet capacity limits.

\section{Ideal: Oracle-based evaluation}
\label{sec:eval}

We evaluate \name in this section and the next section. In this section, we assume that we have a ground truth oracle that gives us the start times, participant locations, and media types of all calls.
We do this to decouple the impact of the prediction error and carve out the utility of \name in oracular settings. \db{Next, in \S\ref{sec:evalcontroller}, we evaluate \name using prediction output.}

\subsection{Metrics of interest}
\label{sec:eval:metrics}

We have four metrics of interest: (a) \emph{Sum of peak network bandwidth (BW) on the WAN links}: Peak network BW (in Tbps) on individual WAN links impacts the network capacity to be provisioned on those links to sustain such peaks. Additionally, peak network BW also drives the network costs\cite{switchboard:sigcomm23}. Thus, we want to lower the sum of peak WAN BW. (b) \emph{Total traffic on WAN links}: Peak network BW does not consider traffic during remaining non-peak times. Thus, we consider total traffic on WAN links across time (peak and non-peak). (c) \emph{E$2$E latency}: user experience depends on the max. E$2$E latency (\S\ref{sec:overview:e2e}). Hence, we evaluate \name using such latencies. (d) \emph{Number of call migrations}: As mentioned in \S\ref{sec:design:realtime}, we may need to migrate the call if the initial assignment of MP DC and routing option is not according to the pre-computed plan. In this section (using a ground truth oracle), the number of migrations is none as the call config is known as apriori. We relax this assumption in the next section, where we assign the MP DC based on the country of the first user.

\subsection{Evaluated policies}
\label{sec:eval:baseline}

In addition to \name, we consider the following three baselines.

\textbf{Weighted Round Robin (WRR):} 
WRR\cite{switchboard:sigcomm23} is easy and practical to implement, and optimizes for compute by balancing calls over multiple MP DCs in the same region. We create buckets for distinct combinations of MP DCs and routing options. Each bucket gets its weight based on its share of compute and the fraction of calls on the Internet. When there are multiple countries in call config, we pick the minimum fraction of calls from its countries. E.g., two DCs have compute capacity as 10K and 20K cores. If the minimum (Internet) fraction for client countries to those DCs are $20\%$ and $15\%$, then we calculate capacities (weights) as $8K$:$2K$:$17K$:$3K$ for buckets. We select \{MP DC, routing option\} bucket using such weights.

\textbf{Locality First (LF):} In this policy\cite{switchboard:sigcomm23}, we assign the MP DC and routing option so as to minimize total latency. We formulate it as a Linear Program (LP). The LP variable is the same as used in \S\ref{sec:lp}. The objective, rather, is to minimize the total latency. The constraints also match the constraints in \S\ref{sec:lp}. We do not include end-to-end latency to avoid it affecting the objective (that anyway considers latency). We also consider a variant that optimizes total max. E2E latency.

\rohan{\textbf{\tilda:}} \tilda selects MP DC through weighted random policy where weights are set in proportion to the number of cores in MP DCs. It then randomly selects calls from individual source country -- destination MP DC pairs based on the capacity calculated in \S\ref{sec:intmigration}.

\textbf{\name (TN):} We use the LP described in \S\ref{sec:lp} to jointly calculate the MP DC and routing option using the oracular ground truth. We use COIN-OR\cite{coinor:web} to run the LP.

Switchboard is not a baseline: (a) It provisions resources months in advance, while \name works within already provisioned limits, and (b) it does not assign routing options or reduce migrations. These are complimentary systems.

\subsection{Data sources}
\label{sec:eval:data}

($1$) \textbf{Latency:} We use the latency between participant and destination countries obtained using our measurements in \S\ref{sec:internet}, ($2$) \textbf{Capacity of the Internet paths:} We estimate the Internet path capacities as derived in \tilda (\S\ref{sec:intmigration}), ($3$) \textbf{Number of calls per call config:} Our telemetry framework logs the number of calls per config. This data is used as the ground truth in this section.

For the evaluation (in this and the next section), we consider all calls that are contained within Europe, \ie where all participants of the calls are in Europe. \tilda has been up and running in Europe for many months with reasonably stable and tested Internet path capacity estimates.

\subsection{Comparing WAN bandwidth}
\label{sec:eval:bw}

\begin{wrapfigure}{l}{0.4\textwidth}
\vspace{-0.2in}
\includegraphics[width=0.4\textwidth, page=2]{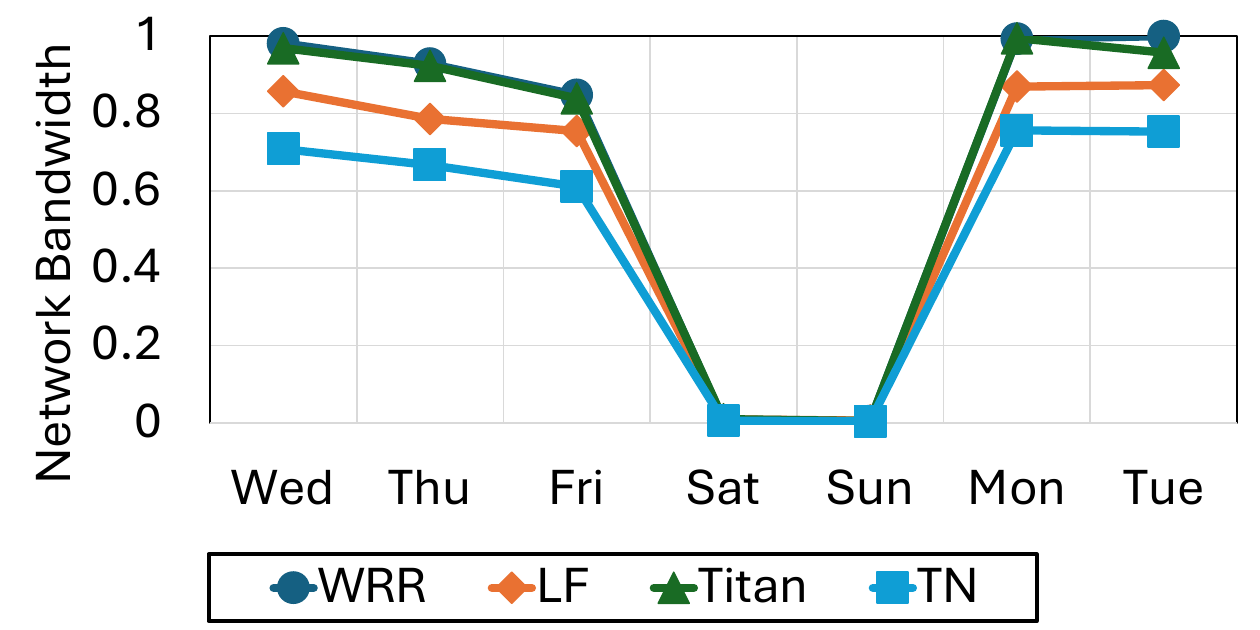}
\caption{Sum of peak bandwidth (BW) on individual links in WAN used by WRR, LF, Titan and \name (TN), calculated each day of the week (\S\ref{sec:eval:bw}). The BW is normalized to peak BW for WRR.}
\label{fig:eval:bw}
\vspace{-0.1in}
\end{wrapfigure}

\textbf{Reduced sum of peaks:} Fig.\ref{fig:eval:bw} shows the WAN sum of peaks bandwidth (BW) for each day of a typical week. On weekdays, when the number of calls is high, \name reduces the WAN BW by $24$-$28\%$ compared to WRR, and $13$-$19\%$ compared to LF. LF places the calls to the MP DCs nearest to users, which in turn reduces the number of hops and WAN BW. However, it is not WAN traffic peak aware. In contrast, TN is peak aware and picks the MP DCs and Internet routes for call configs intelligently. The savings in TN are dominated by current limit on Internet offload (max. 20\%). In reality, some countries had 5-15\% traffic on Internet due to performance deterioration. Moreover, some countries are yet to be flighted to Internet offload in Europe.

\textbf{Savings with only MP DC placement:} In this experiment, we do not assign any calls to Internet for TN only to evaluate the benefits from intelligently assignning MP DCs only. On weekdays, savings in TN compared to WRR and LF reduced to 16.7-20\% and 3-8\%. The remaining savings in TN compared to previous experiment are due to Internet offload. 

\textbf{More savings with more traffic on the Internet:} As mentioned in \S\ref{sec:intmigration}, we are so far conservative in calculating traffic on the Internet. In this experiment, we evaluate the savings with TN if we were to hypothetically double the traffic on the Internet. In such cases, we observed that TN reduces the sum of peak bandwidth by $27$-$38\%$ and $17$-$26.5\%$ compared to WRR and LF respectively (weekdays).

\textbf{Total WAN traffic reduction:} We see similar savings for total WAN traffic (in PB) in TN with Internet offload calculated by \tilda. On weekdays, TN cuts the total traffic by $24$-$28\%$ and $13.5$-$18\%$ compared to RR and LF respectively.

\begin{wraptable}{l}{0.45\textwidth}
\vspace{-0.2in}
\caption{Daily average of max. E2E latency across calls (in msec) for WRR, LF, and \name.}
\small
\begin{tabular}{|c|c|c|c|}
\hline
 & Mean & Median & P$95$ \\
\hline
WRR & $82$ - $86$ & $75$ - $78$ & $120$\\
\hline
LF & $71$ - $75$ & $70$ & $100$ - $103$\\
\hline
\name & $74$ - $80$ & $70$ - $76$ & $103$ - $122$ \\
\hline
\end{tabular}
\label{tab:eval:e2e} 
\vspace{-0.1in}
\end{wraptable}

\textbf{LF using E2E latency:} We consider a variant of LF minimizing total max. E2E latency across configs. TN reduces peak bandwidth against such a policy by 16-29\% (weekdays). 

While we discuss results for a specific week here, our broad observations hold true across weeks.

\subsection{Comparing end-to-end (E2E) latency}
\label{sec:eval:latency}

Table \ref{tab:eval:e2e} shows the daily average of max. E$2$E latencies across all calls for all three policies. WRR is not optimized for latency. LF is specifically optimized for total latency. In contrast, TN is optimized for network bandwidth (BW) (sum of peak BW on individual links) with a constraint on average of max. E2E latency ($E = 80$ in Fig.\ref{fig:algo:lp} for weekends and $E = 75$ for weekdays). Despite TN not being optimized for latency, it achieves latency better than WRR and slightly worse than LF. In TN, we observed infeasible solutions below those values of $E$, while the network BW savings were roughly the same for all values above those values of $E$. This shows that TN can significantly reduce WAN BW compared to LF with a small permissible penalty in E$2$E latency.

\section{Practical: Prediction-based evaluation}
\label{sec:evalcontroller}

In this section, we do not assume ground truth information. We assign the MP DCs and routing option \textit{using country of the first joiner} as detailed in \S\ref{sec:design:realtime}. The \name controller makes such assignment using offline precomputed plan using the prediction output. 

\subsection{Evaluated policies}
\label{sec:evalcontroller:baselines}

We cannot use the WRR, \tilda and LF versions from the previous section, as they assume knowledge of the ground truth. Hence, we modify the baselines to select the MP DC and routing option based on the location of the first user. We evaluate: \textbf{(1) WRR:} We create the buckets so that each bucket has a distinct combination of MP DC and routing option. We assign the weights to the bucket in the same way as described in \S\ref{sec:eval:baseline}, but they are based on the \textit{country of the first user}. \textbf{(2) LF:} 
We sort the MP DC and routing option buckets in descending order based on the latency from the country of the first joiner and pick the first bucket with enough capacity. \textbf{(3) \tilda:} We pick the MP DC and routing option bucket using weighted random based on the country of the first joiner. \textbf{(4) \name (TN):} We use the TN controller that performs real-time assignments using a precomputed plan. We train the Holt-Winters time-series prediction model with $4$ weeks of data to predict the number of calls for individual call configs for the next $24$ hours at the granularity of $30$~min. We feed the prediction output to the LP. We assign the MP DC and routing option as described in \S\ref{sec:design:realtime}. 
\begin{wrapfigure}{l}{0.4\textwidth}
\includegraphics[width=0.4\textwidth, page=3]{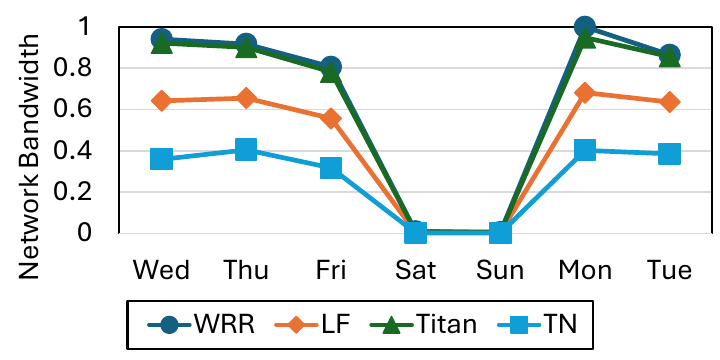}
\caption{Sum of peak bandwidth (BW) on individual links in WAN used by WRR, LF, and \name (TN) calculated for each day of the week (\S\ref{sec:evalcontroller:wan}). The BW is normalized to the peak BW observed for WRR.}
\label{fig:evalcontroller:bw}
\vspace{-0.2in}
\end{wrapfigure}

\subsection{Comparing WAN bandwidth}
\label{sec:evalcontroller:wan}

Fig.\ref{fig:evalcontroller:bw} shows the sum of peak bandwidth (BW) on WAN. TN reduces such BW by $55$-$61\%$ and $38$-$44\%$ on average compared to WRR and LF respectively. The key reason is that LF and WRR no longer have prior knowledge of call configs; they do not know the future call demand. Thus, some of the calls arriving early take the preferred slots while the later calls are assigned far away. TN is peak-aware and uses flexibility in picking the MP DCs and routing options using call history.

\subsection{Accuracy of prediction}
\label{sec:evalcontroller:prediction}

Fig.\ref{fig:evalcontroller:prediction} (Appendix \S\ref{sec:app:accuracy}) shows the RMSE (Root Mean Square Error) and MAE (Mean Absolute Error) in prediction using the Holt-Winters method. Recall that we predict the number of calls for call configs (not reduced call configs). We measure the error for each call config, normalize it to the peak values, and plot the CDF. This way elephant and mice call configs are treated equally. The median errors are small -- $4.9\%$ and $10.6\%$ for MAE and RMSE. 

\subsection{Reduction in migrations}
\label{sec:evalcontroller:migrations}

\begin{wraptable}{l}{0.5\textwidth}
\small
\caption{Percentage of calls that need to be migrated.}
\begin{tabular}{|c|c|}
\hline
With call config & With reduced call config\\
        \hline
        $11$-$34\%$ (average = 31\%) & $11$-$19\%$ (average = 15\%)\\
        \hline
\end{tabular}
\label{tab:evalcontroller:migrations}
\vspace{-0.2in}
\end{wraptable}

As mentioned in \S\ref{sec:design:realtime}, we may migrate the call due to differences in the MP DC and routing option assignments few minutes into the call. As discussed in \S\ref{sec:lp}, we can reduce the number of migrations by using reduced call configs. We compare the volume of call migration needed with and without this approach -- in the former case we feed the reduced call config to the offline LP, while in the latter case, we feed the call config as-is. We only consider inter-DC migrations (not due to routing option changes) as they are more damaging. Table \ref{tab:evalcontroller:migrations} shows that reduced call configs cut down the migrations by $38$-$66\%$ on weekdays (when the number of calls is high). We leave reducing migrations further to future work.

\subsection{\name overhead}
\label{sec:evalcontroller:overhead}

The prediction block runs once a day, grouping and LP blocks run every 30 mins, and the controller runs for each call. The prediction building block takes $1.2$ -- $4.7$~seconds per call config. The entire prediction finishes in $\sim${}$82$~min on a single core. The prediction pipeline is embarrassingly parallel and could span multiple cores to scale, as needed. The call config grouping is very fast (finishes in under a minute). The LP takes roughly $1$ min. Lastly, the controller is again very fast and assigns the MP DC and routing option within $1$~msec per call.

\section{Related work}
\label{sec:related}

\textbf{Conferencing:} Conferencing services such as Zoom\cite{zoom:web}, Microsoft Teams\cite{teams:web}, Google Meet\cite{meet:web}, DingTalk\cite{dingtalk:web}, and others have received considerable community attention\cite{macmillan:imc21, chang:imc21,  carlucci:mmsys16, via:sigcomm16, xron:sigcomm23}. Some of the recent work include: (a) resource management\cite{switchboard:sigcomm23}, (b) network condition based video quality adaptation\cite{gso:sigcomm22}, (c) low latency video transport network\cite{livenet:sigcomm22, lowlatency:sigcomm22}, and (d) codec and transport collaboration\cite{salsify:nsdi18, zhou:mcn19}. In contrast, \name focuses on intelligently reducing costs for these services by leveraging the Internet.

\textbf{The Internet vs. WAN:} Prior works\cite{wan:infocom20, xron:sigcomm23} have analyzed performance of Internet and WAN. Compared to \cite{wan:infocom20}, we have more DCs ($11$ vs. $21$) and significantly more latency measurements ($88K$ vs. $3.5M$) spanning almost the entire globe ($241K+$ cities). 
Unlike \cite{wan:infocom20}, we also measure network loss.
\cite{xron:sigcomm23} analyzes network performance data for source-destination pairs from $11$ DC locations for $1$ day. \cite{fb:internet:imc19, google:internet:imc21, congestion:sigcomm18, slow:internet:pam17} shed light on Internet performance. \cite{li2010cloudcmp} focuses on comparing different cloud providers. \name is orthogonal -- it compares performance metrics for WAN vs. Internet at a significantly larger scale, and uses the insights to select modalities for \teams traffic. 

\textbf{Leveraging multiple communication paths:} \cite{xron:sigcomm23, converge:sigcomm23} are  impressive works that (like \name) use both WAN and Internet paths. \cite{mptcp:ton18, mptcp:wireless21} focus on consuming multiple paths using MPTCP. Such works are complimentary to \name; they do not consider MP server and routing joint assignment. cISP\cite{cisp:nsdi22} uses free-space speed-of-light radio connectivity. SCION\cite{scion:ieee18} argues for path-aware routing. \name leaves it as future work.

\textbf{Traffic engineering (TE):} Many of the TE solutions (\cite{b4:sigcomm13, swan:sigcomm13, contracting:nsdi21, hose:sigcomm21, b4after:sigcomm18}) work for WAN. \cite{espresso:sigcomm17, edgefabric:sigcomm17} focus on TE for Internet. Such works do not have  flexibility of choosing the end-points (MP DCs).

\textbf{Server (MP) selection:} Like MP selection problem in \name, prior work to study server/DC/replica selection\cite{taiji:sosp19, fastroute:nsdi15, donar:sigcomm10, liu:ton15, zhang:ieee13, kwon:cloud14, rubik:atc15}. \cite{c3:nsdi15,shithil:cloud20} focus on replica selection to improve the tail latency. In contrast, \name focuses on joint MP and routing option selection.

\section{Conclusion}
\label{sec:conc}

It is important for large conferencing services like \teams to continue offering a good user experience at low costs. In this paper, through large-scale measurements spanning almost the entire globe ($241K+$ cities), we show that Internet paths provide similar or better latencies in many parts of the world. We present: (a) \tilda (running in production) that calculates the \teams traffic that could be \emph{safely} offloaded to Internet using latency measurements, and (b) \name (research prototype) that jointly assigns the server location and routing option to \teams calls. Together they cut down the WAN sum-of-peak bandwidth, which determines the network cost, by up to $61\%$.

\bibliographystyle{abbrv}
\bibliography{reference}

\appendix


\section{Internet and WAN performance}
\label{sec:app:1}

\subsection{Loss on Internet and WAN}
\label{sec:app:loss}

In this experiment, we measure the number of $30$~min time-slots  over a $7$~day period where the loss on individual paths (Internet or WAN) is at least $0.1\%$. Fig.\ref{fig:app:loss} shows the CDF for all ($176$) client country - MP DC pairs in Europe. It can be seen that Internet has more frequent loss. $50\%$ source - MP DC pairs suffer loss of at least $0.1\%$ on Internet for at least $2\%$ time-slots. In contrast, $0.1\%$ loss on WAN is rare -- the number of time-slots at $P100$ is bound to $2\%$.

We repeat the same analysis when the loss is minimum $1\%$. As expected, there are fewer time-slots when the loss on Internet is $\geq${}$1\%$ versus when loss is $\geq${}$0.1\%$. However, even in this case, Internet still has more frequent loss compared to WAN.

\subsection{Elasticity on Internet}
\label{sec:app:elasticity}

\begin{figure*}
\centering
    \begin{minipage}{.5\textwidth}
        \centering
        \includegraphics[width=0.95\textwidth, page=1]{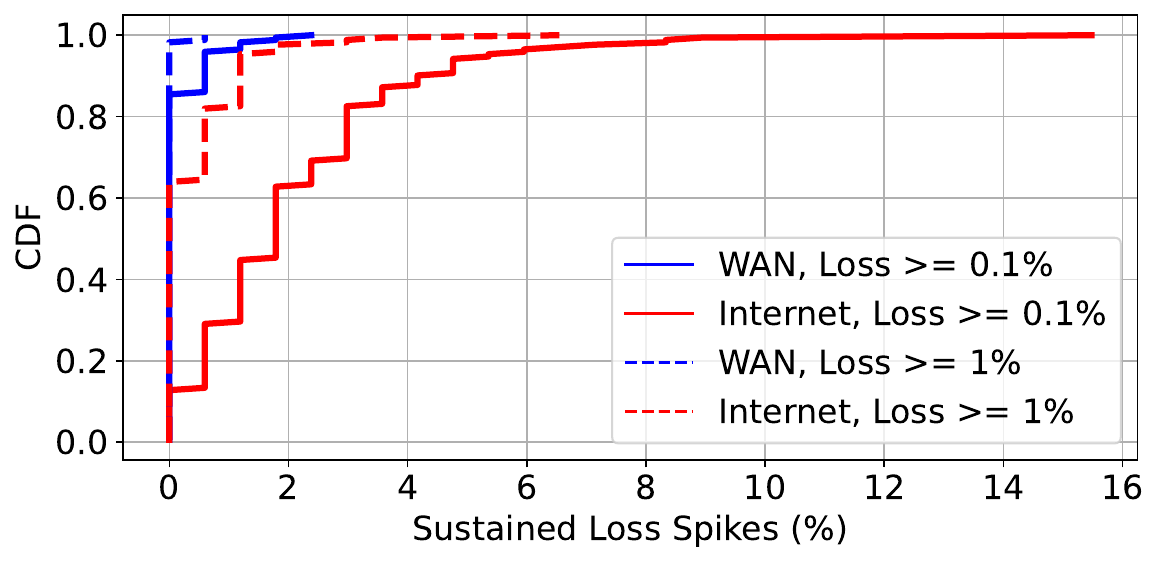}
        \caption{CDF of packet loss across all client countries - MP DC pairs in Europe.}
        \label{fig:app:loss}
    \end{minipage}%
    \hspace{0.1cm}
    \begin{minipage}{0.46\textwidth}
        \centering
        \includegraphics[width=0.95\textwidth, page=1]{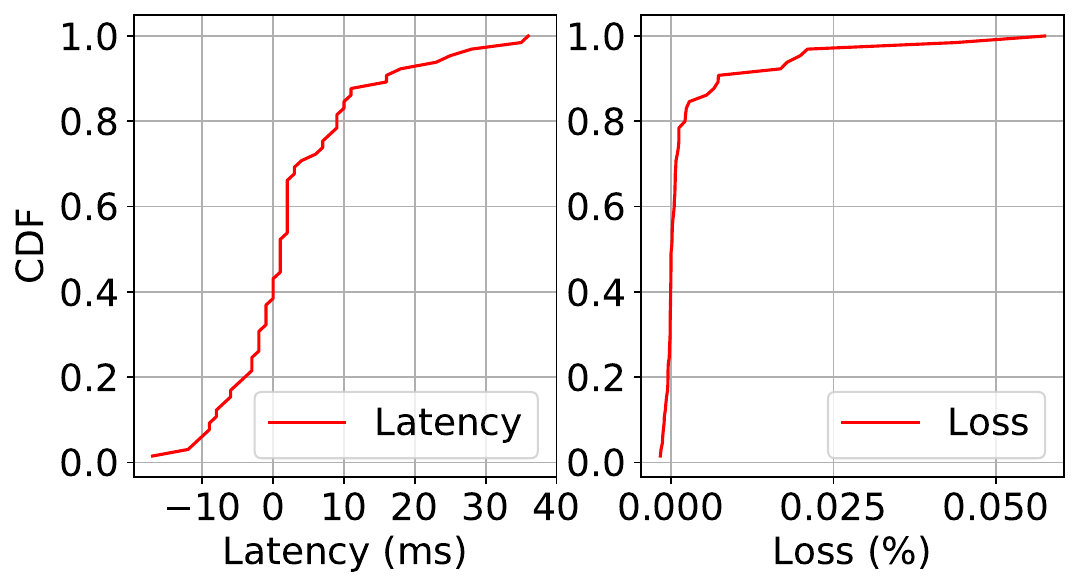}
        \caption{Elasticity on Internet.}
        \label{fig:app:elasticity}
    \end{minipage}%
\end{figure*}

Fig.\ref{fig:app:elasticity} shows the increase in latency and loss between different client country - MP DC pairs in Europe as we increase the fraction of traffic on Internet from $1\%$ to $20\%$. Note that the changes in latency and loss are impacted by two factors: (a) traffic moved by \tilda from $1\%$ to $20\%$, and (b) underlying infrastructure and routing changes outside \tilda. Often, it is not possible to decouple these factors as \tilda takes a few months (multiple trial-and-error) to complete movement of traffic to Internet, and it has no visibility into the ISP infrastructure. The negative latency difference is likely because Internet infrastructure improved over a period of time. It can be seen that the latency difference even at $P90$ is under $20$~msec. Even for loss, the difference at $P90$ is under $0.01\%$. Internet providers either have capacity available or are able to quickly add capacity as needed. These results show that Internet is fairly elastic to accommodate increasing traffic from \tilda. 

\begin{SCfigure}[1][h]
\centering
\includegraphics[width=0.45\textwidth, page=1]{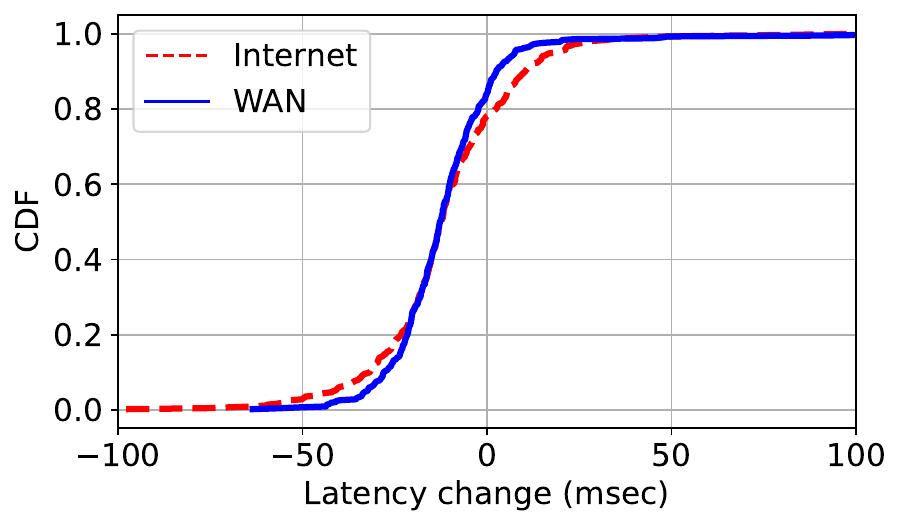}
\caption{Latency difference between 2 weeks separated by 12 months. Negative values = improvements.}
\label{fig:measure:trend}
\end{SCfigure}

\subsection{Long-term trends:} 
\label{sec:app:trend}

For both the Internet and WAN paths between the $20$ countries (top; by call volume) and all DCs, we measured the weekly median latencies for the weeks that are $\sim${}$12$~months apart. Fig.\ref{fig:measure:trend} plots the CDFs of changes in latency (new minus old; negative means improvement) for the Internet and WAN paths. While for more than $80\%$ cases latencies have improved for both types of paths, the Internet paths see slightly greater improvements.

\subsection{Impact of fine grained granularities}
\label{sec:app:gran}

Fig.\ref{fig:measure:latencyheatmap} shows the fraction ($F$) of times (when considering hourly median values for $1$~week) Internet paths offer latencies lower than or comparable ($\leq${}$10$~ms inflation) to WAN paths from different source countries to destination DCs. To do such an analysis, we consider the clients at the granularity of a \textit{country}. However, one client country can have different ASNs or cities with potentially different performance (consequently different $F$). In this section, we detail the difference in $F$ when considering different granularities compared to granularity of country as shown in Fig.\ref{fig:measure:clarity}. Let's consider granularity of city + ASN. Let's say one country has $N$ combinations of city + ASN, with fractions $F$ as $\{F_{1}$ to $F_{N}\}$. The fraction $F$ for that country is $F_{c}$. Similarly, the fractions of number of measurements for individual combination of city + ASN to total number of measurements for that country are $\{w_{1}$ to $w_{N}\}$ (${\displaystyle\sum_{i \in N}{w_{i}} = 1}$). For each city + ASN combination, we calculate the difference in $F$ compared to granularity of country as follows: The difference ($D$) is calculated as $D = \frac{\displaystyle\sum_{i \in N}{|F_{i} - F_{c}|} \cdot w_{i}}{F_{c}}$. We calculate $D$ for each client country - destination MP country and plot $P50$ and $P90$ in Fig.\ref{fig:measure:clarity}. It can be seen that difference $D$ is bound to $11\%$ even at $P90$ for city + ASN. These results show that granularity of country performs similar to more fine grained granularities such as ASN and city + ASN.


\subsection{Stability}
\label{sec:app:stability}

\begin{figure*}[t]
\centering
\includegraphics[width=0.9\textwidth]{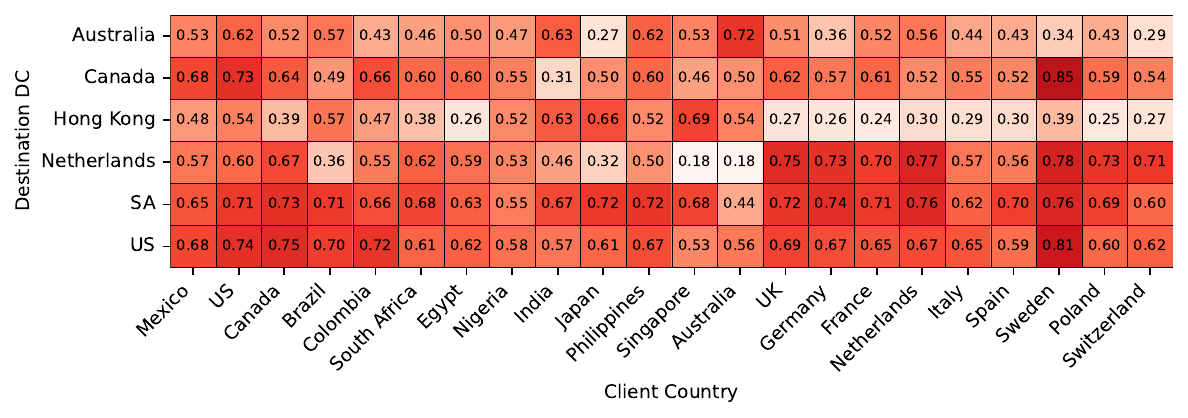}
\vspace{0.05in}
\caption{Fraction of times Internet provides better or comparable (within $10$~msec) latency compared to WAN. We show for different source countries and $6$ \azure DCs. SA denotes South Africa and US denotes the United States. We use $1$-week data from the month of December'$23$ that is $6$ months apart from the dates used in Fig.\ref{fig:measure:latencyheatmap}.}
\label{fig:app:1}
\end{figure*}

We perform the same analysis as in Fig.\ref{fig:measure:latencyheatmap} using data from a week in December'$23$ ($6$ months apart). The results are shown in Fig.\ref{fig:app:1}. We find that, in general, the trends are similar as described in Fig.\ref{fig:measure:latencyheatmap}, while the North America - Europe corridor has improved slightly in $6$ months.

\section{Accuracy of prediction}
\label{sec:app:accuracy}

\begin{wrapfigure}{l}{0.4\textwidth}
\centering
\includegraphics[width=0.4\textwidth, page=1]{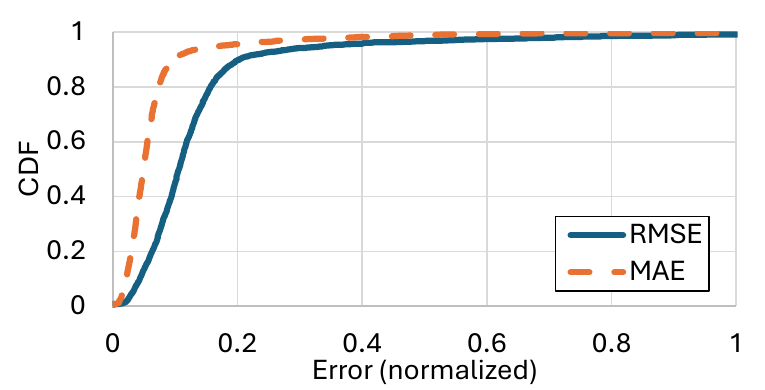}
\caption{CDF of error (normalized to max values).}
\label{fig:evalcontroller:prediction}
\vspace{-0.2in}
\end{wrapfigure}

Fig.\ref{fig:evalcontroller:prediction} shows the CDF for accuracy of prediction in terms of RMSE (Root Mean Square Error) and
MAE (Mean Absolute Error) across $3$,$000$ call configs. The Holt-Winters based prediction in \name is fairly accurate with median errors of $4.9\%$ and  $10.6\%$ for MAE and RMSE respectively. $95.6\%$ ($89.7\%$) call configs have normalized MAE (RMSE) less than $20\%$. For a small number of call configs, the errors are relatively large due to unexpected change in the number of calls for such configs.

\end{document}